\documentclass[fleqn,usenatbib,useAMS]{mnras}

\usepackage{graphicx}	
\usepackage{amsmath}	
\usepackage{amssymb}	
\usepackage{multicol}        
\usepackage{bm}		
\usepackage{pdflscape}	
\usepackage{multirow}
\usepackage{ulem}

\newcommand{\target}{J191213.72$-$441045.1} 
\newcommand{\targ}{J1912$-$4410} 
\newcommand{\eff}{_{\textrm{\tiny eff}}} 

\usepackage[T1]{fontenc}
\usepackage{ae,aecompl}

\title[The white dwarf in \targ]{Unveiling the white dwarf in \target\ through ultraviolet observations}

\author[Pelisoli et al.]{
Ingrid Pelisoli$^{1}$\thanks{E-mail: ingrid.pelisoli@warwick.ac.uk}, Snehalata Sahu$^{1}$, Maxim Lyutikov$^{2}$, Maxim Barkov$^{3}$, Boris~T. G\"{a}nsicke$^{1}$,
\newauthor Jaco Brink$^{4,5}$, David A.~H. Buckley$^{4,5,6}$, Stephen. B. Potter$^{4,7}$, Axel Schwope$^{8}$, S.~H. Ram\'{i}rez$^{1}$
\\
$^{1}$Department of Physics, University of Warwick, Gibbet Hill Road, Coventry, CV4 7AL, UK\\
$^{2}$Department of Physics and Astronomy, Purdue University, 525 Northwestern Avenue, West Lafayette, IN, USA\\
$^{3}$Institute of Astronomy, Russian Academy of Sciences, Moscow, 119017 Russia\\
$^{4}$South African Astronomical Observatory, PO Box 9, Observatory, 7935, Cape Town, South Africa\\
$^{5}$Department of Astronomy, University of Cape Town, Private Bag X3, Rondebosch 7701, South Africa\\
$^{6}$Department of Physics, University of the Free State, PO Box 339, Bloemfontein 9300, South Africa\\
$^{7}$Department of Physics, University of Johannesburg, PO Box 524, Auckland Park 2006, South Africa\\
$^{8}$Leibniz-Institut f\"{u}r Astrophysik Potsdam (AIP), An der Sternwarte 16, 14482 Potsdam, Germany
}
\date{Last updated XXX; in original form XX}

\pubyear{2023}

\begin{document}
\label{firstpage}
\pagerange{\pageref{firstpage}--\pageref{lastpage}}
\maketitle

\begin{abstract}
\target\ is a binary system composed of a white dwarf and an M-dwarf in a 4.03-hour orbit. It shows emission in radio, optical, and X-ray, all modulated at the white dwarf spin period of 5.3~min, as well as various orbital sideband frequencies. Like in the prototype of the class of radio-pulsing white dwarfs, AR~Scorpii, the observed pulsed emission seems to be driven by the binary interaction. In this work, we present an analysis of far-ultraviolet spectra obtained with the Cosmic Origins Spectrograph at the {\it Hubble Space Telescope}, in which we directly detect the white dwarf in \target. We find that the white dwarf has a temperature of $T\eff = 11485\pm90$~K and mass of $0.59\pm0.05$~M$_{\sun}$. We place a tentative upper limit on the magnetic field of $\approx50$~MG. If the white dwarf is in thermal equilibrium, its physical parameters would imply that crystallisation has not started in the core of the white dwarf. Alternatively, the effective temperature could have been affected by compressional heating, indicating a past phase of accretion. The relatively low upper limit to the magnetic field and potential lack of crystallisation that could generate a strong field pose challenges to pulsar-like models for the system and give preference to propeller models with a low magnetic field. We also develop a geometric model of the binary interaction which explains many salient features of the system.
\end{abstract}




\section{Introduction}

Binary white dwarf pulsars are systems composed of a fast spinning white dwarf and a late-type main sequence star that show strong pulsed emission on the white dwarf spin period, detectable from radio to X-rays \citep{Marsh2016, Pelisoli2023}. Their broad-band luminosity cannot be explained by the stellar components alone, nor by any accretion mechanisms: they have low X-ray luminosities, display no aperiodic broad-band variability \citep[also referred to as flickering, e.g.][]{Scaringi2014} characteristic of accreting systems, and typically only show narrow emission lines, indicating that no significant accretion occurs \citep[e.g.][]{Garnavich2019}.
Unlike the canonical neutron star pulsars, it is believed that the source of emission is intrinsically tied to binarity and is due to magnetic interaction between the two stars: free electrons are accelerated to near relativistic speeds as the magnetic field of the white dwarf sweeps past the companion, generating non-thermal pulsed synchrotron emission \citep{Geng2016, Takata2017, Katz2017, Lyutikov2020}.

The prototype of this class is AR~Scorpii \citep[AR~Sco,][]{Marsh2016}, which was serendipitously discovered after being misclassified for many years as a $\delta$-Scuti pulsating star, due to the orbital modulation resembling the saw-tooth shape shown by the light curves of radial pulsators. In reality, the observed 3.56-hour modulation is also seen in the radial velocity of the M-dwarf companion, indicating a binary origin. The asymmetric shape can likely be attributed to phase-dependent contribution from non-thermal emission \citep{Katz2017}. As well as the orbital modulation, high-speed optical photometry revealed strong pulses with a period of 1.97 minutes, interpreted as the beat period between white dwarf spin and orbit. The pulses were subsequently detected also in radio \citep{Stanway2018} and X-rays \citep{Takata2018}. These strong pulses allow for precise timing of the white dwarf spin, which was found to be slowing down at a high rate of $P/\dot{P} = 5.6 \times 10^6$~years \citep{Gaibor2020, Pelisoli2022c}. Additionally, strongly pulsed ($\sim 90\%$ pulse fraction) linear polarisation, of up to $40\%$, was also detected \citep{Buckley2017}, primarily modulated at the 1.95 min spin period. This was consistent with a strong dipole field ($\gtrsim 200$ MG), exhibiting beamed synchrotron emission within its magnetosphere \citep{PotterBuckley2018}. The strong  magnetic field was inferred from the assumption that the luminosity is dominated by synchrotron emission from a rotation-powered dipole \citep{Marsh2016, Buckley2017}.




The combination of a fast spin, suggestive of an initially low magnetic field ($\lesssim 10$~MG), allowing the white dwarf to accrete and gain angular momentum, and rapid spin-down, pointing at a high magnetic field ($\gtrsim 100$~MG) capable of providing a synchronising torque, made of AR~Sco a challenge to models of accreting binaries. Two main classes of models were put forward trying to reconcile AR~Sco's puzzling observed characteristics:

\begin{enumerate}
\item {\it High magnetic field}: \citet{Katz2017} proposed that the rapid spin-down could be explained by magnetic torque, which would require the white dwarf and M-dwarf to have magnetic fields of $\sim 100$~MG and $\sim 100$~G, respectively. In this scenario, the spin-down power is dissipated in the atmosphere of the M-dwarf by magnetic reconnection, which produces the observed synchrotron radiation. Similar magnetic field strengths were assumed by \citet{Geng2016} and \citet{Takata2017} in their modelling of AR~Sco's pulse profile and spectral energy distribution. The downside of these models is that such a high magnetic field would prevent the white dwarf from accreting enough matter to explain its present spin period --- a very large mass transfer rate of up to $\dot{M} \sim 10^{-4}\,\mathrm{M}_\odot \,\mathrm{yr}^{-1}$ \citep{Ghosh1979, Lyutikov2020} would be required to compress the magnetosphere for enough accretion to occur. Such a rate is at least $10^5$ times greater than estimated values for similar binaries \citep{Pala2022}. Additionally, there is to date no direct detection of AR~Sco's magnetic field. \citet{Garnavich2021} constrained it to $B \lesssim 100$~MG based on the lack of Zeeman splitting of the Lyman-$\alpha$ line.

\item {\it High mass-transfer rate:} An alternative model was proposed by \citet{Lyutikov2020}, who suggested that the magnetic field cannot be larger than $\sim 10$~MG for the white dwarf to have been spun up to current rates. They assumed a more typical mass transfer rate of $\dot{M} \sim 10^{-9}\,\mathrm{M}_\odot \,\mathrm{yr}^{-1}$ and argued that, if the ionisation rate in the M-dwarf wind is not high, neutral particles will travel through the magnetic field lines of the white dwarf unaffected. Close enough to the white dwarf, they are exposed to ultraviolet radiation and ionised, couple to the magnetic field and are then centrifugally expelled from the system, carrying away angular momentum. This would imply that AR~Sco is in a propeller state similar to AE~Aquarii \citep[e.g.][]{Patterson1979, Chincarini1981, Eracleous1996} and LAMOST~J024048.51+195226.9 \citep{Thorstensen2020, Garnavich2021b, Pretorius2021, Pelisoli2022a}. However, unlike the confirmed propellers, AR~Sco shows no observational evidence of flaring, which argues against a propeller behaviour.
\end{enumerate}

A recent model for the evolution of magnetic white dwarfs in close binary stars proposed by \citet{Schreiber2021} could potentially reconcile a fast spinning white dwarf and a high magnetic field without the need for unfeasible mass transfer rates. They proposed that the white dwarfs in magnetic cataclysmic variables were not born magnetic, allowing for unimpeded accretion-driven spin-up, and only became magnetic due to a rotation- and crystallisation-driven dynamo \citep{Isern2017}. This proposition resolves the theoretical issue with spinning up a highly magnetic white dwarf, reinstating models with a significant magnetic field as a possibility. In short, the two proposed classes of models remain possible, but with observational shortcomings in both cases: there is no detection of a high magnetic field in AR~Sco, arguing against models requiring a significant field, but there is also no strong evidence of flaring, contradicting models proposing a significant mass-transfer rate.

The discovery of a second binary white dwarf pulsar, \target\ (henceforth \targ), by \citet{Pelisoli2023} and \citet{Schwope2023} provided the first opportunity to test the theoretical models put forward to explain AR~Sco. Like the prototype of the class, \targ\ contains a compact object and an M-dwarf in a close binary. The orbital period is 4.03~h and the spin period, which in this case dominates over the beat, is 5.3~min and is also detected from radio to X-rays. The length of the spin period, over a factor of four longer than any confirmed neutron star pulsar \citep{Caleb2022}, led to the interpretation of the system as a second binary white dwarf pulsar. Due to its recent discovery, the spin-down of the white dwarf has not been constrained yet, but the spectral energy distribution shows excess flux compared to the stellar components, in a similar manner to AR~Sco, pointing at possible spin-down power.

The current theoretical framework makes three observables key in determining the feasibility of proposed theoretical models: i) the white dwarf magnetic field, ii) the white dwarf temperature (which determines the level of crystallisation), and iii) the mass transfer rate. \citet{Pelisoli2023} reported potential flaring in \targ, which could suggest a significant mass-transfer rate, but pointed out that continuous monitoring is required to confirm their findings, in particular as the flares could potentially be attributed to the M-dwarf companion rather than to a propeller behaviour. The white dwarf magnetic field and its temperature, on the other hand, could not be determined as the optical emission is completely dominated by the irradiated face of the M-dwarf; only an upper limit of $T\eff < 13\,000$~K was estimated. This work fills this gap by analysing far-ultraviolet (FUV) {\it Hubble Space Telescope} ({\it HST}) observations of \targ, which reveal the spectrum of the white dwarf.

\section{Observations and data reduction}

\targ\ was observed during Cycle 30 for 12 orbits, split into three visits of four orbits each. Exposure details of each visit are shown in Table~\ref{tab:visits}. For both visits 2 and 3, failure in guide-star acquisition prevented science exposures for one of the four orbits. We used the Cosmic Origins Spectrograph (COS) instrument with the G140L grating centred at 1105~\AA, providing flux-calibrated coverage between $\approx1110$ and $\approx2150$~\AA. TIME-TAG mode was employed, where the position and detection time of every photon is recorded with 32~ms precision in so-called {\tt corrtag} files.

\begin{table}
	\centering
	\caption{Details of each {\it HST} visit to \targ.}
	\label{tab:visits}
	\begin{tabular}{cccc} 
		\hline
		  Visit & Orbit  & Exposure start (UTC) & Exposure duration (s) \\
		\hline
        \\
		  \multirow{4}{*}{1} & 1 & 2023-03-18 21:24:46 & 2508.192 \\
		& 2 & 2023-03-18 22:54:34 & 2782.144 \\
		& 3 & 2023-03-19 00:29:45 & 2782.144 \\
		& 4 & 2023-03-19 02:04:57 & 2782.144 \\  
        \\
        \multirow{4}{*}{2} & 5 & 2023-05-05 15:22:14 & 2508.160 \\  
        & 6 & \multicolumn{2}{c}{no data acquired} \\ 
        & 7 & 2023-05-05 18:28:47 & 2782.176 \\  
        & 8 & 2023-05-05 20:07:37 & 2782.176 \\  
        \\
        \multirow{4}{*}{3} & 9 & 2023-05-06 15:12:58 & 2508.128 \\  
        & 10 & \multicolumn{2}{c}{no data acquired} \\
        & 11 & 2023-05-06 18:18:17 & 2782.176 \\  
        & 12 & 2023-05-06 19:53:40 & 2782.144 \\  
        \\
		\hline
	\end{tabular}
\end{table}

\begin{figure}
	\includegraphics[width=\columnwidth]{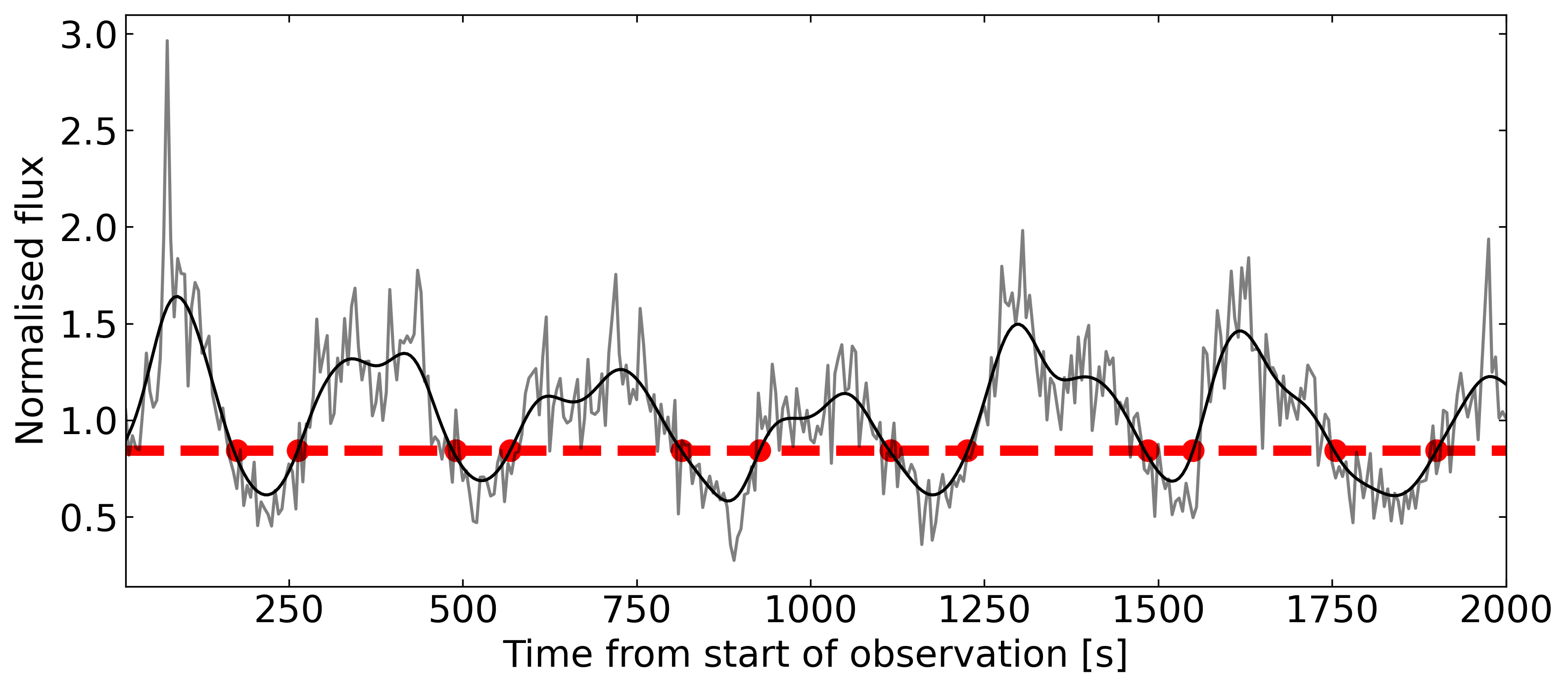}
         \caption{A section of \targ's light curve, obtained during orbit 8, illustrating how the boundary of pulses was defined. The grey line shows the normalised light curve and the black line shows its smoothed version. The red dashed line marks the 30\% percentile. The crossing points between the smoothed data and this line were assumed as the change between peak and off-peak status.
         }
    \label{fig:pulses}
\end{figure}

We downloaded all {\tt corrtag} files and used the {\sc lightcurve} package\footnote{\url{https://github.com/justincely/lightcurve/}} to obtain a light curve from the observations, masking airglow\footnote{\url{https://www.stsci.edu/hst/instrumentation/cos/calibration/airglow}} features and limiting the wavelength range to 1100 to 2000~\AA\ as preliminary inspection showed reduced signal-to-noise above 2000~\AA. Following the approach that \citet{Garnavich2021} employed for AR~Sco, we use the light curve to identify the times of emission between the pulses, where the contribution from non-thermal emission should be at a minimum and the white dwarf can be revealed. Our approach was to first normalise the light curve to minimise the orbital modulation. To do this, we smoothed the light curve by convolving it with a Gaussian kernel with a width given by $\sigma = 50$~s, which dilutes the pulse contribution, and then divided the original light curve by the smoothed version. Next we smoothed the normalised light curve using a 5~s Gaussian kernel, to remove oscillations caused by noise. The boundary times between off-pulse and pulse contributions was then determined as the 30 per cent quantile, as illustrated in Fig.~\ref{fig:pulses}.

New {\tt corrtag} files were generated for peak and off-peak exposures using the {\sc splittag} function from the {\sc costools} package\footnote{\url{https://github.com/spacetelescope/costools}}. Spectra were then extracted for each file using the COS pipeline {\sc calcos} (version 3.4.6) and downloaded reference files (version hst\_1080.pmap). All contributions identified as between or during a peak were averaged to create the off-peak and peak spectra, respectively.

A similar procedure was applied to create orbital phase-resolved spectra. We created {\tt corrtag} files and extracted spectra for ten orbital phase bins of equal size (0.1). Finally, we also combined the two approaches to extract off-peak spectra only around orbital phase 0.5 (0.4--0.6), when the white dwarf is at inferior conjunction.

\section{Photometric analysis}

\begin{figure}
	\includegraphics[width=\columnwidth]{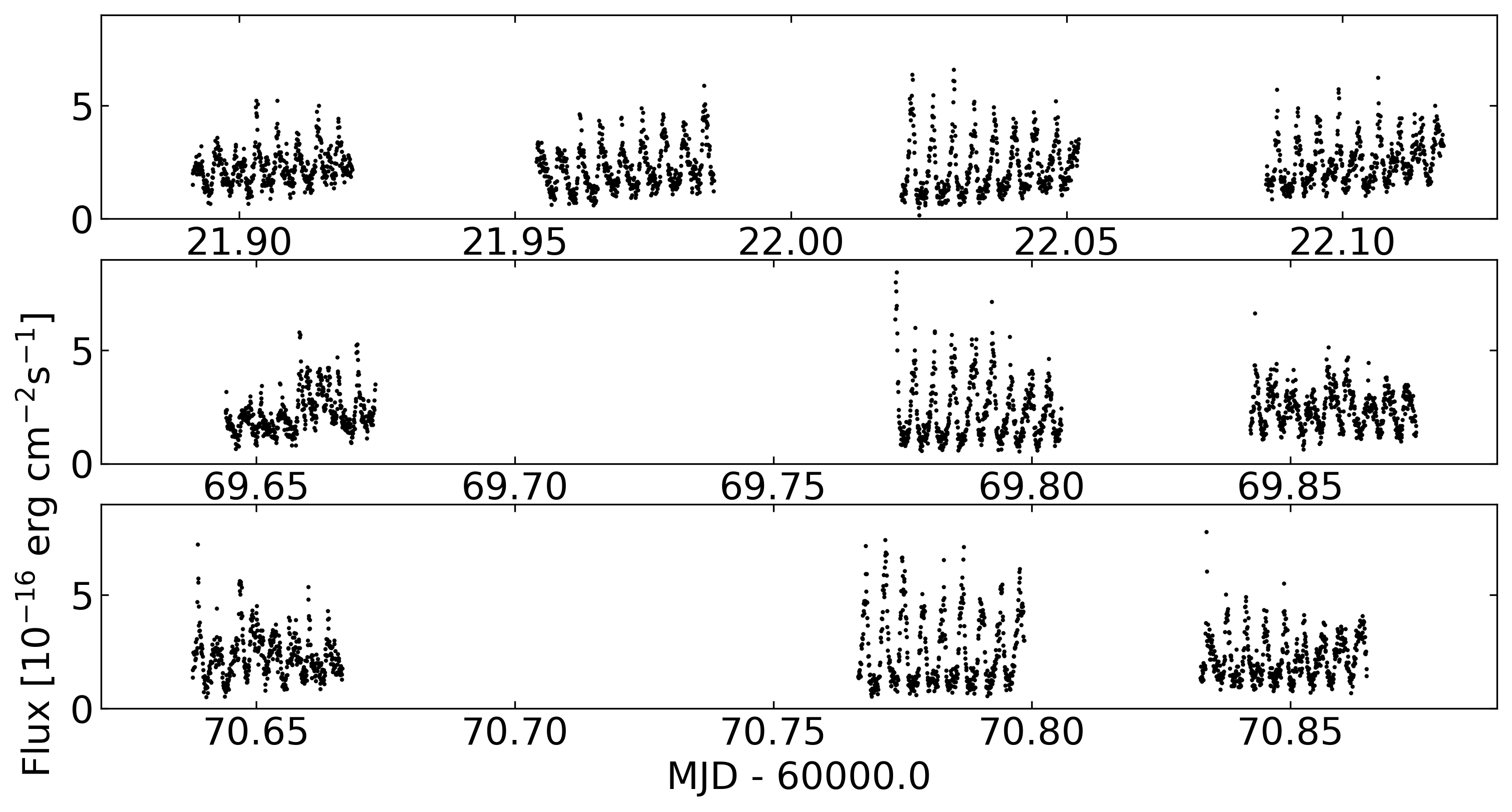}
    \includegraphics[width=\columnwidth]{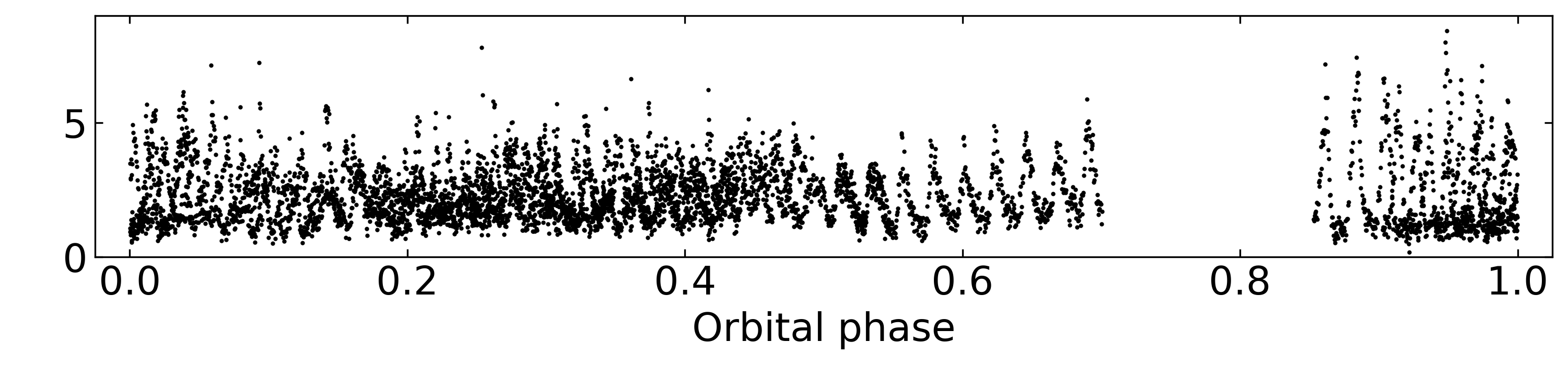}
         \caption{The top three panels show the light curve of \targ\ for each of the three {\it HST} visits, with flux integrated between 1100 and 2000~\AA. The gaps between data correspond to different orbits; the large gaps in the middle panels are due to failure in data acquisition during orbits 8 and 10. The bottom panel shows all data folded to the orbital ephemeris of \citet{Pelisoli2023} to illustrate our phase coverage.
         }
    \label{fig:lc}
\end{figure}

\begin{figure}
	\includegraphics[width=\columnwidth]{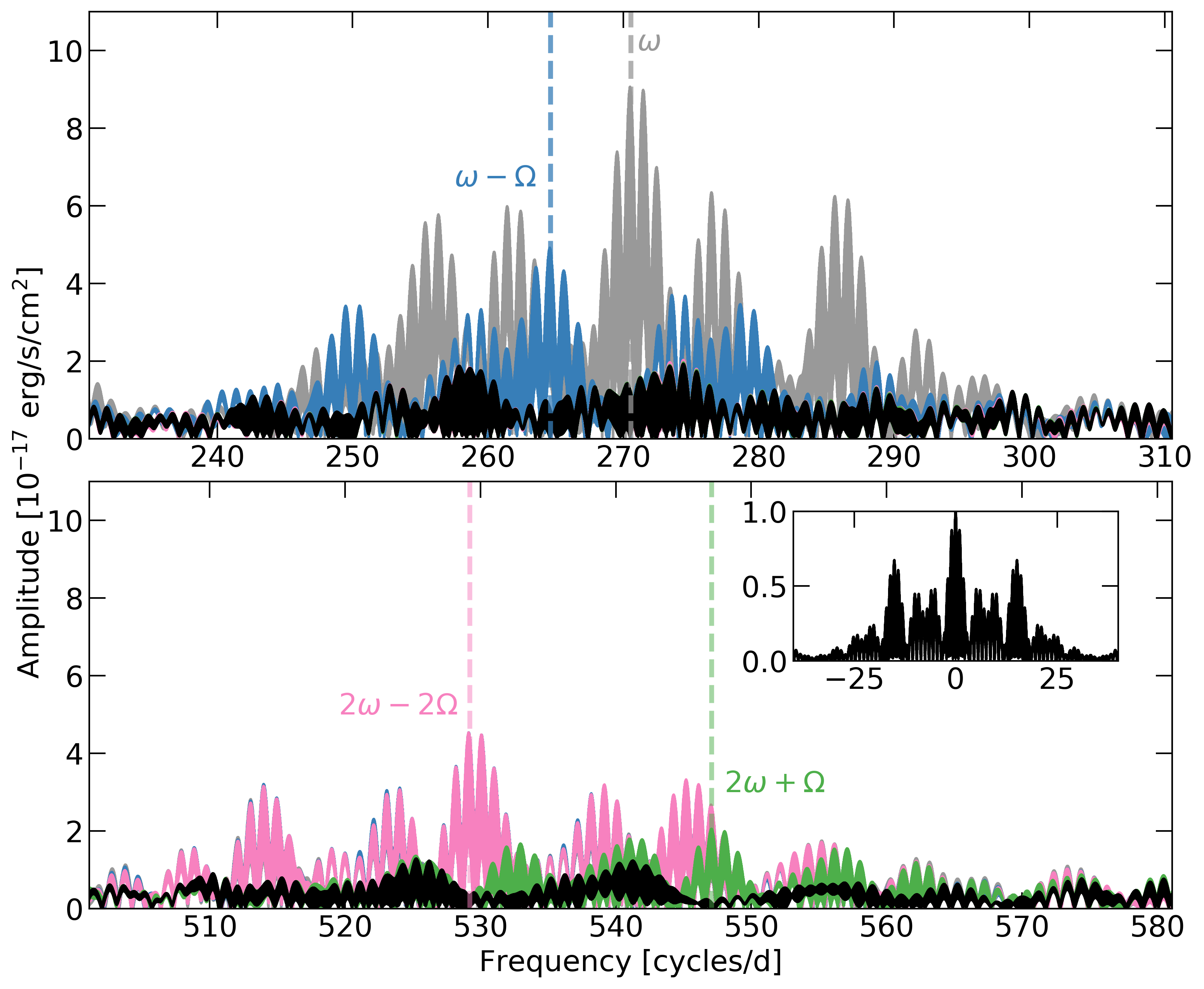}
         \caption{Fourier transform of the ultraviolet light curve of \targ\ around the spin frequency (top panel) and its first harmonic (bottom panel). The inset in the bottom panel shows the window function at the cadence of the {\it HST} observations. There is strong aliasing due to \targ's orbital visibility. The different colours represent different pre-whitening stages, with the dominant frequency at each stage marked by a dashed vertical line of the same colour. The subtracted frequencies were, in order, $\omega$ (spin), $\omega - \Omega$ (beat), $2\omega - 2\Omega$ (beat's first harmonic), and $2\omega + \Omega$.
         }
    \label{fig:ft}
\end{figure}

The obtained FUV light curve for \targ\ is shown in Fig.~\ref{fig:lc}. Like for the radio, optical and X-ray data, the pulses are evident in the FUV data. The flux increases by up to a factor of $\approx 8$ in a few tens of seconds. The Fourier transform of the data is shown in Fig.~\ref{fig:ft}. The dominant frequency is consistent with the one interpreted by \citet{Pelisoli2023} as the spin $\omega$ of the white dwarf. The next strongest contribution is, however, from the {\it beat} frequency $\omega - \Omega$ (where $\Omega$ is the orbital frequency), which was remarkably undetectable in the optical light curves. The next significant contributions around the spin and its first harmonic are from the beat's first harmonic (also detected in the optical) and from $2\omega + \Omega$.

\begin{figure}
	\includegraphics[width=\columnwidth]{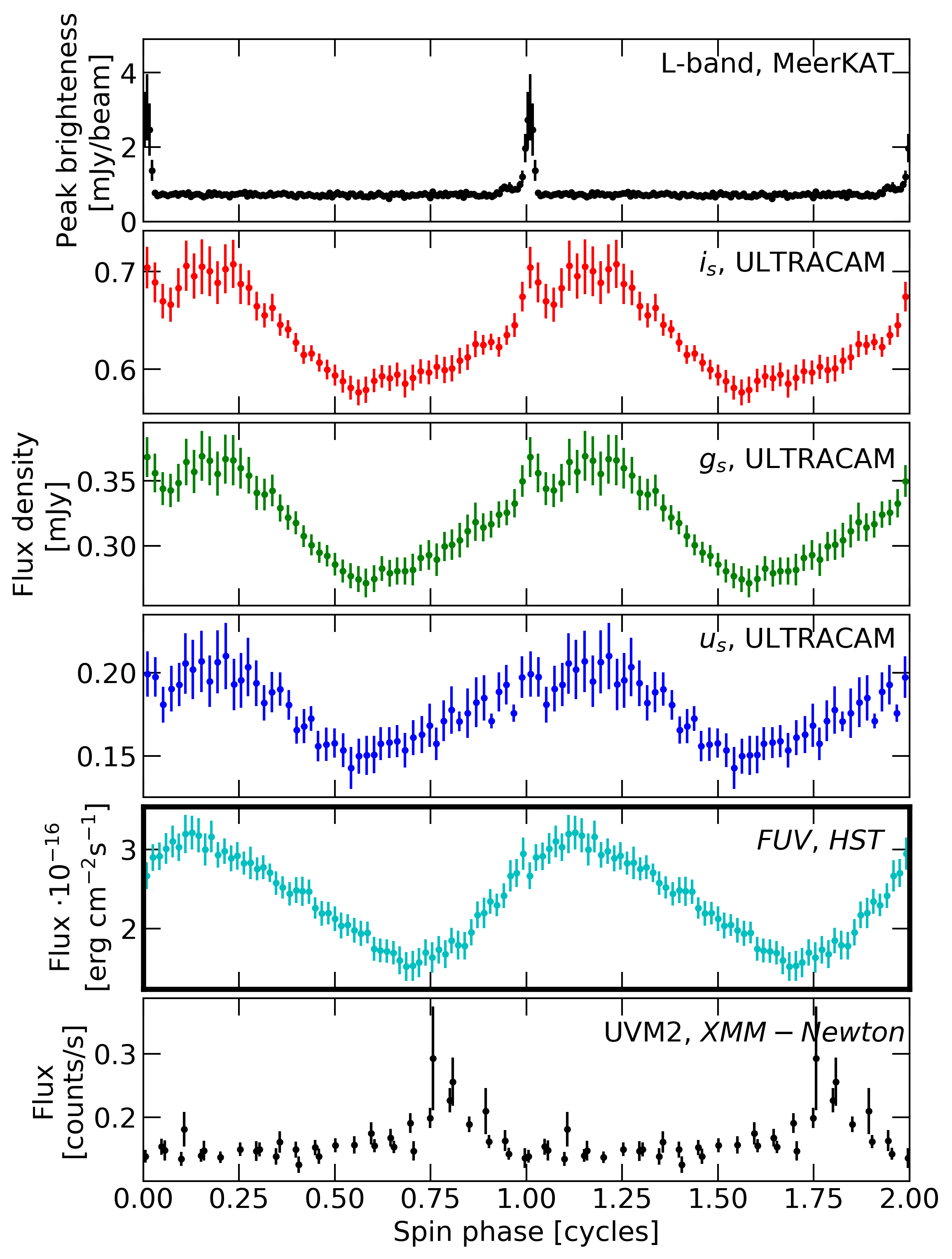}
         \caption{Data in the multiple available bands folded to the spin ephemeris. Radio (top), ULTRACAM (second, third and fourth from the top), and X-ray data (bottom) are from \citet{Pelisoli2023}. The far-FUV data (second panel from the bottom, with the thicker border) were obtained as part of this work.
         }
    \label{fig:sphase}
\end{figure}

Fig.~\ref{fig:sphase} shows the FUV light curve folded to the spin ephemeris reported by \citet{Pelisoli2023}, that is $BJD(TDB) = 2459772.142522(24) + 0.0036961693(10)E$, where $E$ is an integer cycle number. Only the data within orbital phases 0.35 to 0.65 are used for the radio, optical and FUV plots, because the pulse shape shows some orbital dependence. As can be seen, the ephemeris seem to apply reasonably well to the FUV data. There is a hint of narrow peak, like the one dominant in radio and seen in the $i_s$, $g_s$, and $u_s$ ULTRACAM bands, near phase 1.0 as expected. The broader pulse peaks shortly after phase 1.0, as also seen in the optical\footnote{Note that the position of the broad peak compared to the narrow peak depends on the adopted ephemeris. \citet{Schwope2023} adopt slightly different ephemeris and find the narrow peak to appear after the broad peak.}. With the adopted ephemeris, the optical broad pulse seems to be $\approx 0.25$ ahead of the narrow pulse in phase, and the X-rays are $\approx 0.25$ behind. The FUV data look somewhat transitional between the optical and the X-rays: the broad pulse peaks between the peak for optical and X-rays, and the pulse shape is a mix between the optical and X-rays. However, we caution that there seems to be a stochastic or at least not fully understood aspect to the behaviour of the pulses. \citet{Pelisoli2023} noted that, in optical data taken simultaneously with X-ray observations, the optical and X-ray broad pulses were aligned \citep[as also noted in][]{Schwope2023}, but these pulses were offset compared to the spin ephemeris, which could describe well longer optical observations taken before and after the X-ray simultaneous data (see their extended data figure 4). In short, the observed offset might not represent a persistent behaviour and could depend on other system parameters.

As the FUV data extends the previous five-month baseline by over nine months, it might be tempting to combine the {\it HST} data with that from \citet{Pelisoli2023} to further refine the spin ephemeris. However, given the strong dependence of the peak location with wavelength, that can unfortunately not be done. We have attempted to do so by following the same procedure as in \citet{Pelisoli2023}, i.e. start from trial ephemeris to define windows around the expected location of peaks, cross-correlate the data in each window with a Gaussian function with width given by a standard deviation of 15~s to determine the times of maxima in each window from the maximum of the cross-correlation function, and repeat the procedure until the trial and fitted ephemeris are consistent. This resulted in a very poor fit of linear ephemeris, likely due to the natural shift between optical and FUV. In fact, the linear fit was so poor that an $F$-test indicated that the addition of a quadratic term would result in a significant improvement with a confidence level of over 99.9 per cent, but the quadratic term was highly dependent on the orbital phases and datasets included in the fit. In short, probing for spin-down cannot be done by combining data taken with significantly different filters, and will need to wait for more optical data extending the previous baseline.

\section{Spectroscopic analysis}

\begin{figure}
	\includegraphics[width=\columnwidth]{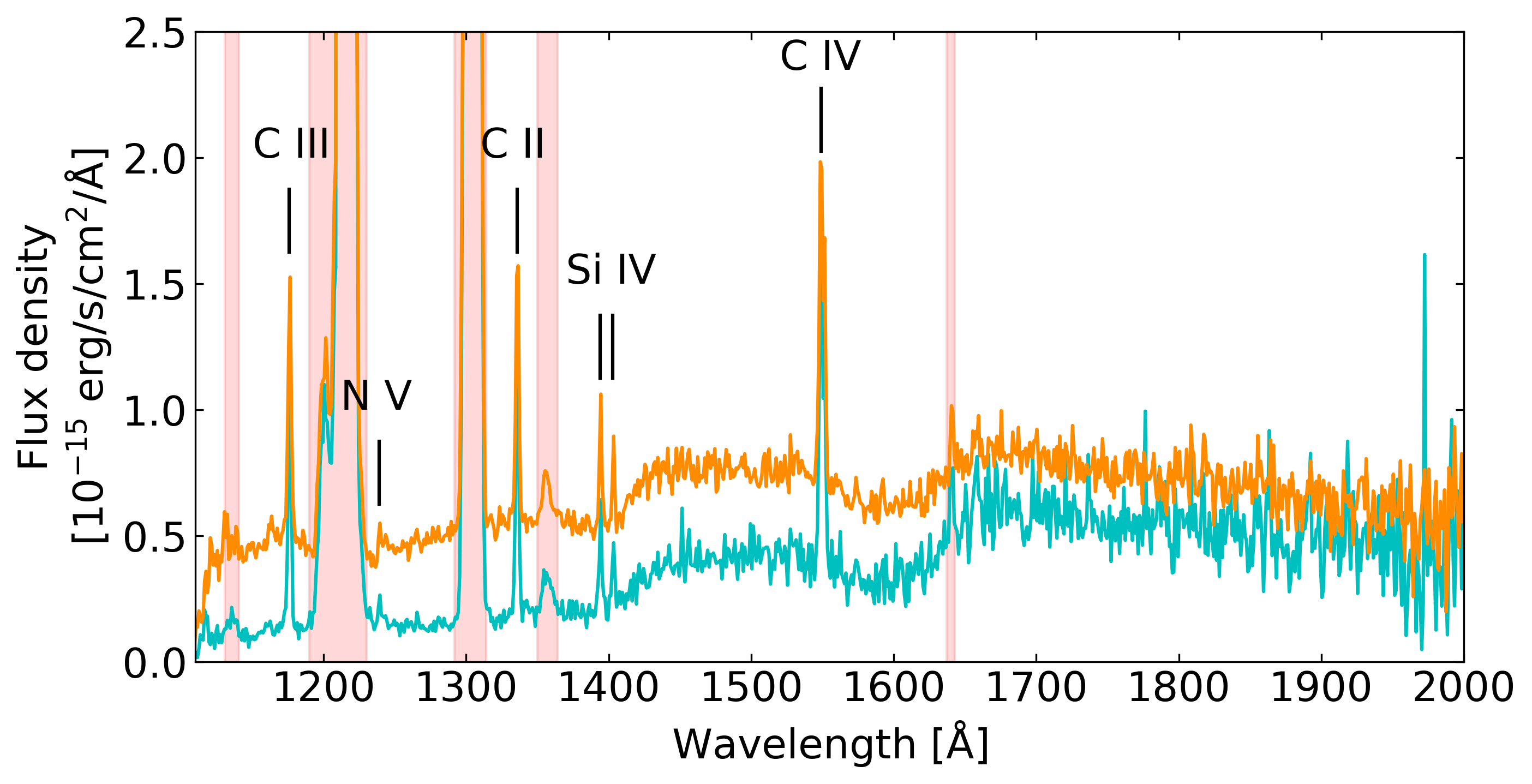}
         \caption{Spectra extracted from the peak (orange, top line) and off-peak (cyan, bottom line) exposures. The regions shaded in red are known airglow features. Some emission lines are likely stellar in origin and can be attributed to the irradiated face of the M-dwarf; these are indicated by the labelled black lines.
         }
    \label{fig:specs}
\end{figure}

\begin{figure}
	\includegraphics[width=\columnwidth]{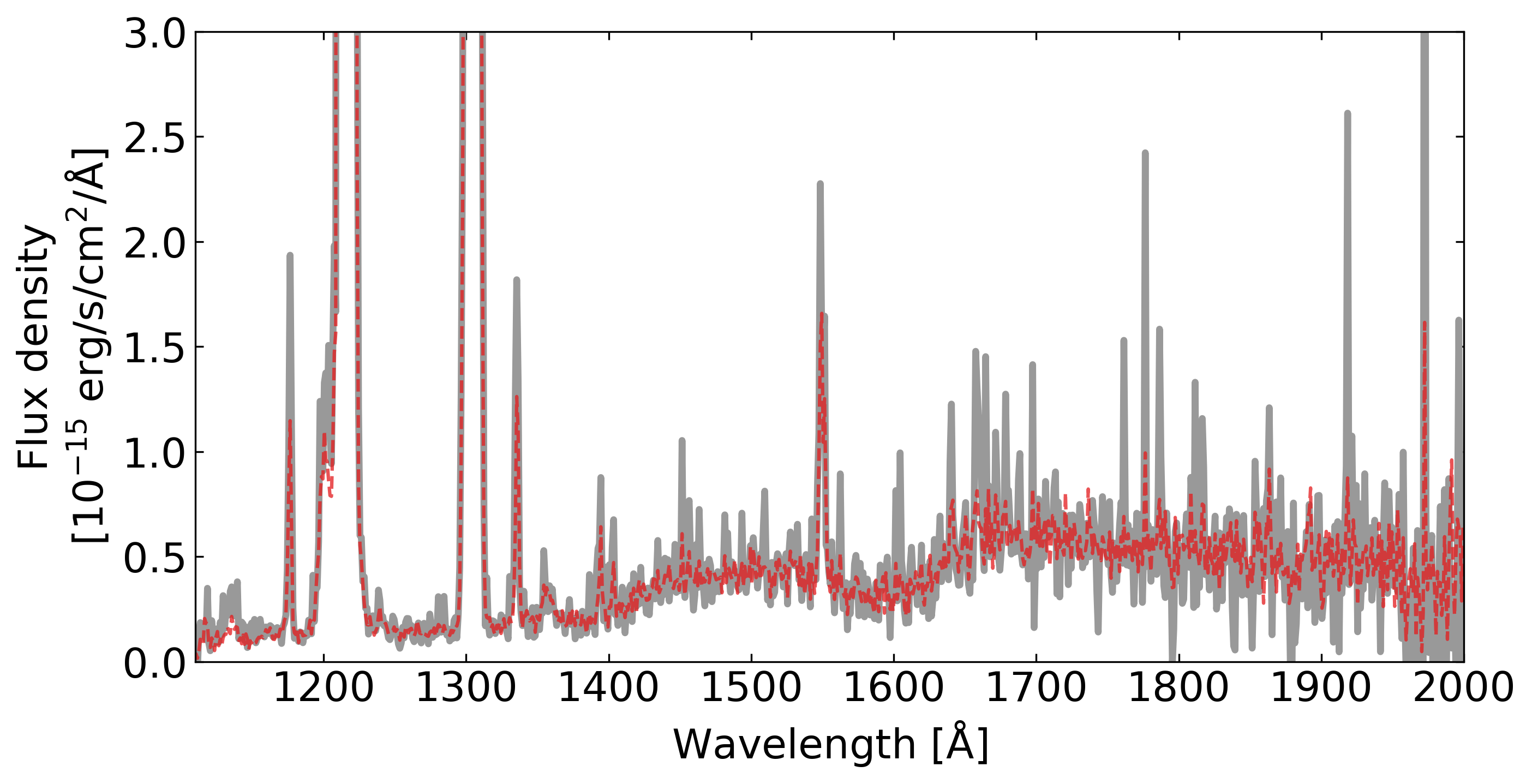}
         \caption{The solid grey line shows the spectrum obtained by extracting data only from the off-peak data between orbital phases 0.4 and 0.6. The red dashed line uses all off-peak data regardless of orbital phase. The difference in flux is seen at the metal emission lines, but Lyman-$\alpha$ and other white dwarf features show no significant change other than decreased noise when all phases are included. 
         }
    \label{fig:comp_specs}
\end{figure}

The obtained peak and off-peak spectra using all orbital phases are shown in Fig.~\ref{fig:specs}. The emission lines are a combination of airglow features and emission lines likely originating on the surface of the irradiated M-dwarf. Features from the white dwarf, in particular the quasi-molecular H$_2$ absorption around 1600~\AA, are visible for both peak (pulse) and off-peak (through) spectra, suggesting a significant contribution from the white dwarf at all spin phases, perhaps due to a favourable inclination. It is also noticeable that, even in the through spectra, there is significant flux at Lyman-$\alpha$, whereas white dwarf models indicate near zero flux. This suggests that there is still some dilution from the pulsed emission. This remains true for a through spectrum extracted only around phase 0.5 (inferior conjunction of the white dwarf), which in fact is completely consistent with the spectrum obtained without any orbital phase constraint (see Fig.~\ref{fig:comp_specs}). Given that the only apparent effect of selecting on orbital phase was to decrease the signal-to-noise ratio, we carried out the analysis of the white dwarf using spectra from all phases.


To estimate the white dwarf spectrum and remove any diluting contribution, our approach followed \citet{Garnavich2021} in subtracting a scaled version of the pulsed emission from the through spectrum. Removing the pulsed emission scaled by 20 per cent resulted in a near-zero flux consistent with white dwarf models. We used the resulting spectrum to estimate the white dwarf physical parameters (Section~\ref{sec:specfit}) and constrain its magnetic field (Section~\ref{sec:bconstrain}). We also analyse the phase-resolved spectra to investigate the orbital dependence of the metal emission lines and derive dynamical mass constraints (Section~\ref{sec:metal_rvs}).

\subsection{Determining the white dwarf's physical parameters}
\label{sec:specfit}

We fit the white dwarf spectrum using white dwarf synthetic spectra computed from an updated grid of pure hydrogen atmosphere models based on \citet{Koester2010}. The updates include the recalculations of unified profiles of Ly$\alpha$ and Ly$\beta$ using new atomic data \citep{Santos2012, Hollands2017} and the use of Stark broadening profiles of \citet{Tremblay2009}. Convection is treated in the mixing length approximation parameterised with $\alpha = 0.8$. Line opacities are included for hydrogen, and molecular opacities for the quasi-molecules H$_2$ and H$_2^{+}$. The fit was performed by $\chi^2$ minimisation, with the parallax fixed at the value reported in {\it Gaia} data release 3 \citep[DR3,][]{gaia1,gaia2,Lindegren2021}, $4.20\pm0.08$~mas, and reddening fixed at the value reported in 3D STILISM models, E(B-V) = 0.035+/0.004 \citep{Lallement2019}. Emission lines were masked as they do not originate on the white dwarf. The parallax allows for a precise radius estimate, which combined with a mass-radius relationship, provides a mass estimate. We used the evolutionary La Plata models with a progenitor metallicity of $Z=0.02$ and a  hydrogen layer with a varying hydrogen mass fraction depending on the white dwarf mass \citep[$\simeq10^{-3}$ for masses less than 0.4~M$_{\sun}$ to $\simeq10^{-6}\,\rm{M_{H}}/M_{WD}$ for 1.1\,M$_{\sun}$,][]{Althaus2013, Camisassa2016, Camisassa2019}.

We carried out three different approaches to the spectroscopic fit: (i) fitting the estimated white dwarf spectrum (obtained by subtracting 20 per cent of the pulse spectrum from the through), (ii) fitting the through spectrum with a white dwarf model and a power law (i.e. modelling the dilution with a power law rather than with the pulse spectrum), and (iii) fitting the through spectrum with a white dwarf model plus dilution from constant flux.

\begin{figure}
	\includegraphics[width=\columnwidth]{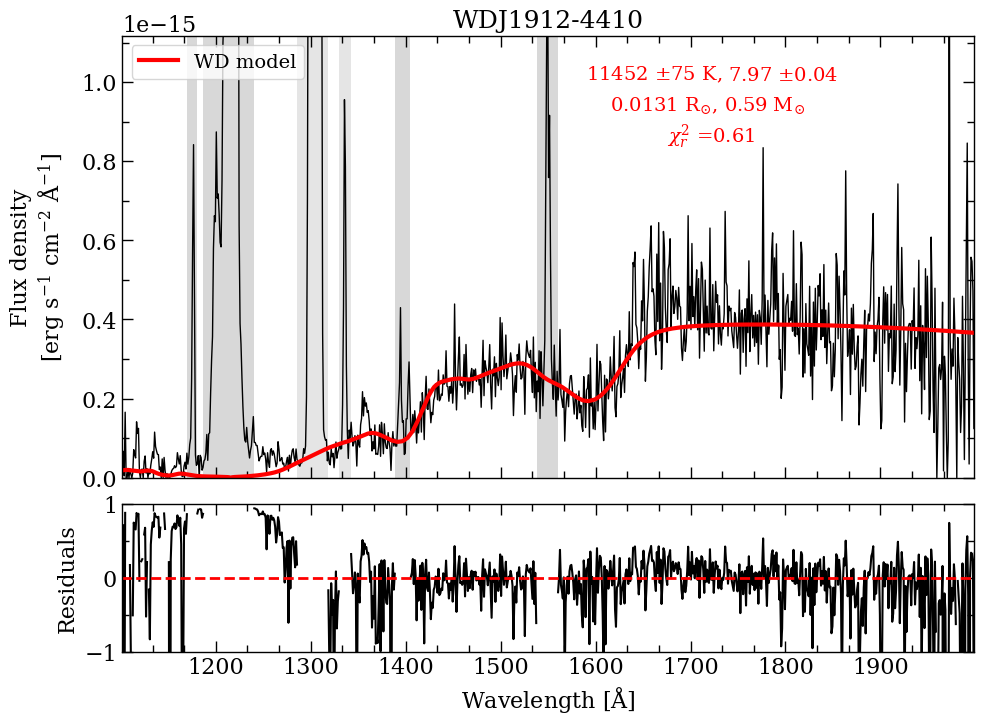}
	\includegraphics[width=\columnwidth]{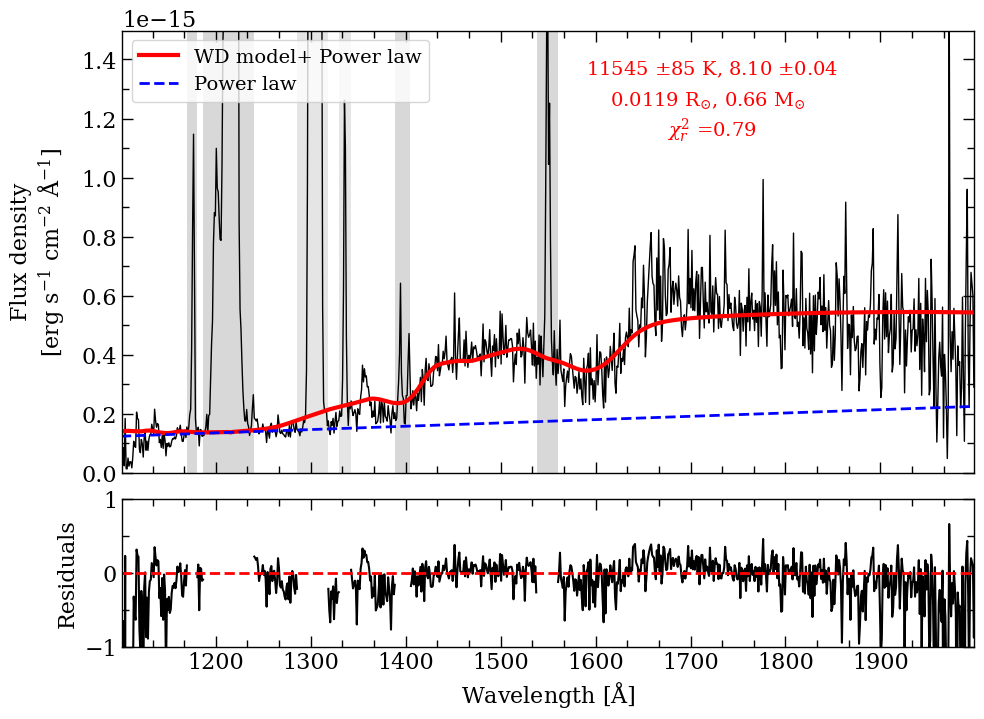}
	\includegraphics[width=\columnwidth]{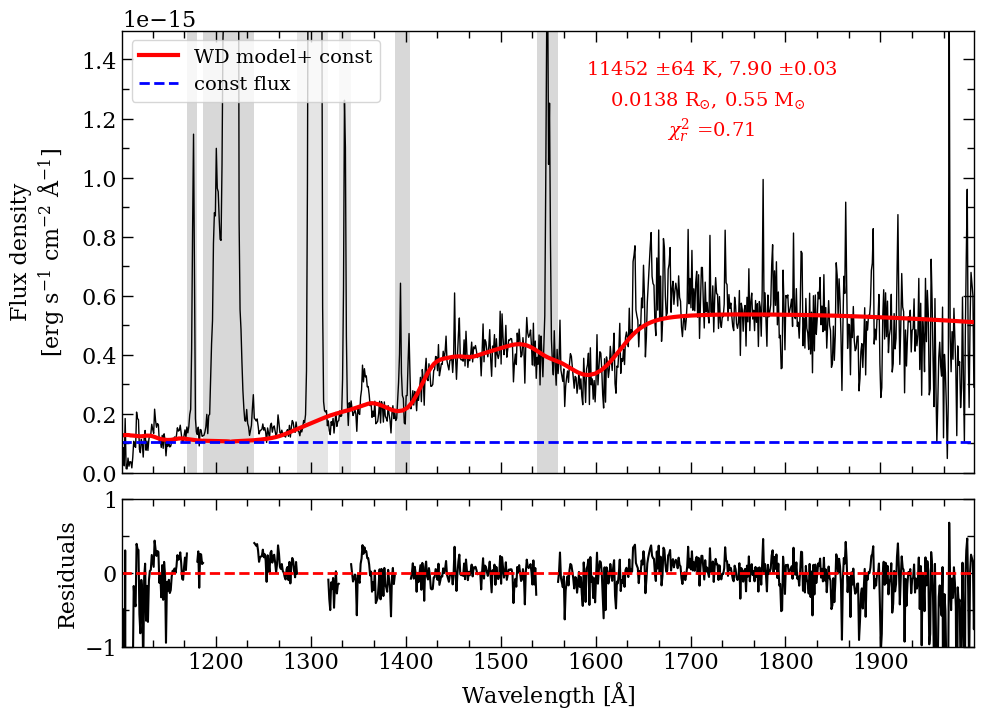}
         \caption{The top panel shows the estimated white dwarf spectrum in black and the best-fit model in red. The emission lines that have been masked for the fit are indicated by the shaded grey areas. The bottom and middle panels show the through spectrum in black, the best model in red, and the model for the additional flux (power law or constant) as a dashed blue line. The best fit values ($T\eff$, $\log~g$, radius, mass and reduced $\chi^2$) are labelled in red in all the figures.
         }
    \label{fig:wdfit}
\end{figure}

The resulting fits are shown in Fig.~\ref{fig:wdfit}. For approach (i), we obtained a temperature of $T\eff{}_1 = 11\,452\pm75$~K and surface gravity with $\log~g_1 = 7.97\pm0.04$. The parallax constrains the radius to $R_1 = 0.0131\pm0.004$~R$_{\sun}$\footnote{We use subscript 1 to refer to white dwarf parameters, and 2 for M-dwarf.}, implying a white dwarf mass of $M_1 = 0.59\pm0.02$~M$_{\sun}$. Similar values are obtained when using approaches (ii) and (iii), as indicated in Table~\ref{tab:fits}. Given that we obtain consistent values between all approaches, we adopt as final parameters the mean between the three methods, with uncertainty given by the standard deviation. This results in $T\eff{}_1 = 11\,485\pm90$~K and $M_1 = 0.59\pm0.05$~M$_{\sun}$. 

\begin{table*}
	\centering
	\caption{White dwarf parameters obtained using the three fitting methods described in the text. The last column reports the reduced $\chi^2$ of each fit. The uncertainties are statistical only. The adopted values are the mean and standard deviation between the three approaches.}
	\label{tab:fits}
	\begin{tabular}{cccccc}
		\hline
        \hline
		  Method & $T\eff$ (K) & $\log~g$ & Radius (R$_{\sun}$) & Mass (M$_{\sun}$) & $\chi_r^2$\\
		\\
        (i) & $11452\pm75$ & $7.97\pm0.04$ & $0.0131\pm0.004$ & $0.59\pm0.02$ & 0.61 \\
        (ii) & $11545\pm85$ & $8.10\pm0.04$ & $0.0119\pm0.004$ & $0.66\pm0.02$ & 0.79 \\
        (iii) & $11452\pm64$ & $7.90\pm0.03$ & $0.0138\pm0.004$ & $0.55\pm0.02$ & 0.71 \\
        \hline
        Adopted & $11485\pm90$ & $8.00\pm0.09$ & $0.128\pm0.040$ & $0.59\pm0.05$  &  \\
		\hline
        \hline
	\end{tabular}
\end{table*}

\subsection{Constraining the magnetic field of the white dwarf}
\label{sec:bconstrain}

\begin{figure}
	\includegraphics[width=\columnwidth]{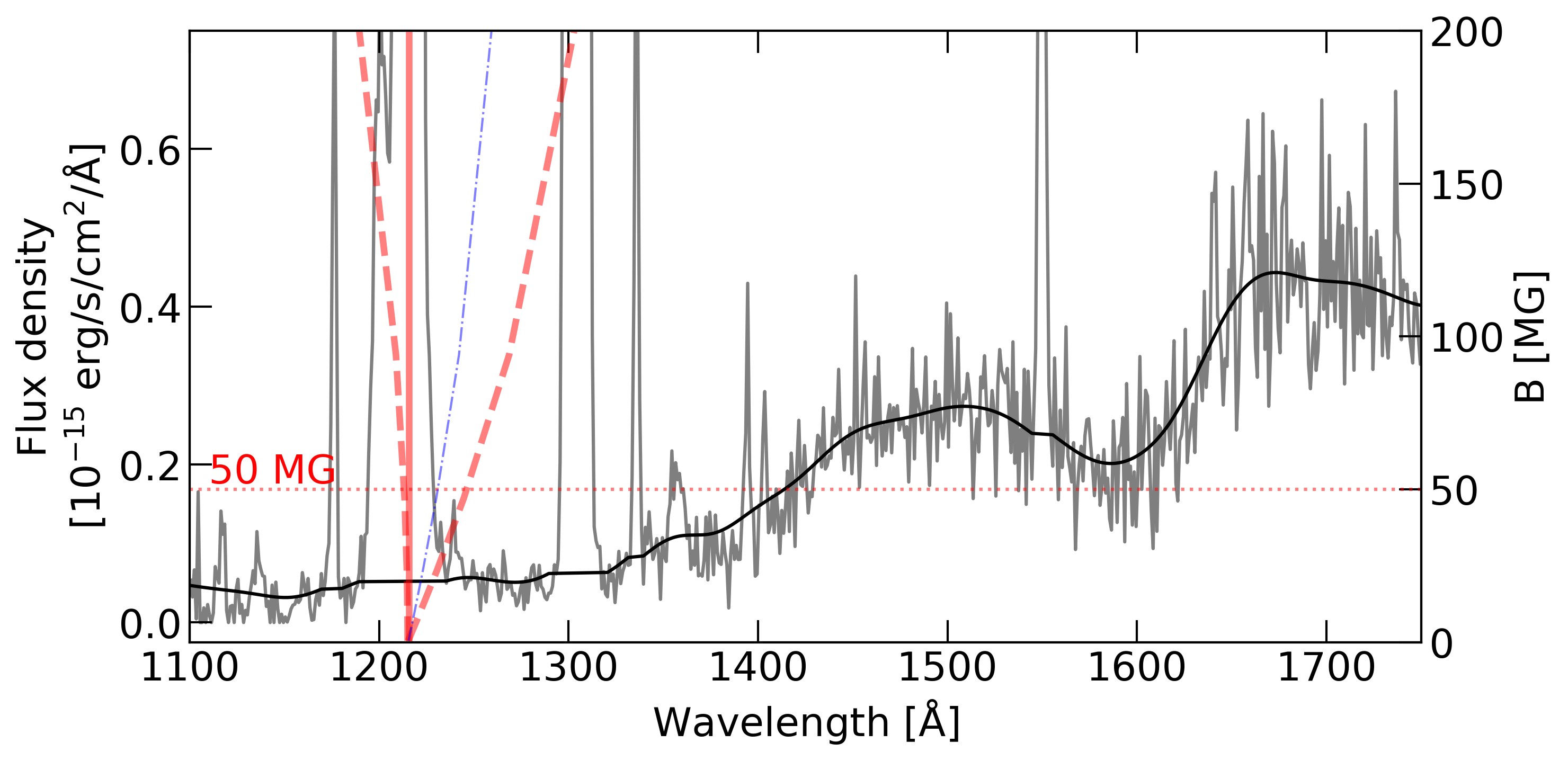}
         \caption{The grey line shows the white dwarf spectrum. The black line is the spectrum smoothed by masking the emission lines and convolving with a Gaussian of standard deviation equal to 5~\AA. The solid vertical red line indicates the rest wavelength of Lyman-$\alpha$, and the dashed red lines show the central wavelength of displaced Zeeman components according to models from \citet{Schimeczek2014} for the magnetic field strength indicated in the right-hand $y$-axis. The blue dot-dashed line shows the midpoint between the central component and the displaced component, that shows the larger shift (on the right). This is where we would expect an increase in flux if there was a magnetic field with a strength larger than $\approx 50$~MG (the dotted horizontal line), as the gap between components would be beyond the geocoronal emission in Lyman-$\alpha$.
         }
    \label{fig:maxb}
\end{figure}

Magnetic fields are detected in white dwarf spectra primarily due to Zeeman splitting of spectral lines. The existence of a magnetic field introduces a preferential direction, which lifts the degeneracy of energy levels on the magnetic quantum number $m$, leading to transitions with different energies for the same spectral line. The separation between line components depends on the strength of the magnetic field, thus allowing for a magnetic field estimate. The spectrum of \targ\ shows, however, no sign of Zeeman splitting, preventing a precise estimate of the white dwarf's magnetic field. The lack of splitting can instead place an upper limit on the magnetic field, above which we would expect to see a sign of different line components. We used the energy levels calculated by \citet{Schimeczek2014} for hydrogen in a magnetic field to calculate the theoretical wavelength for the Zeeman split components of the $1s$ to $2p$ transition, which corresponds to Lyman-$\alpha$. As illustrated in Fig.~\ref{fig:maxb}, up to a magnetic field of $\approx 50$~MG, the gap between components would be filled by the strong geocoronal emission centred in Lyman-$\alpha$. Above $\approx 50$~MG, one of the gaps would be beyond the geocoronal emission and should be detected as an increase in flux. As this is not detected, we estimate that the magnetic field of the white dwarf in \targ\ is $\lesssim 50$~MG.

Another way to detect magnetic fields via spectroscopy, in particular for accreting systems, is to identify humps corresponding to the cyclotron frequency of electrons accelerated by the magnetic field \citep[e.g.][]{Schwope2006}. We would expect cyclotron humps to be detected in the FUV for magnetic fields $\sim 100-200$~MG, hence the non-detection here makes fields in this range unlikely, though cyclotron humps for stronger fields would appear in the near-UV, beyond our wavelength coverage.

\subsection{Inferring dynamics from the metal emission lines}
\label{sec:metal_rvs}

\begin{figure}
	\includegraphics[width=\columnwidth]{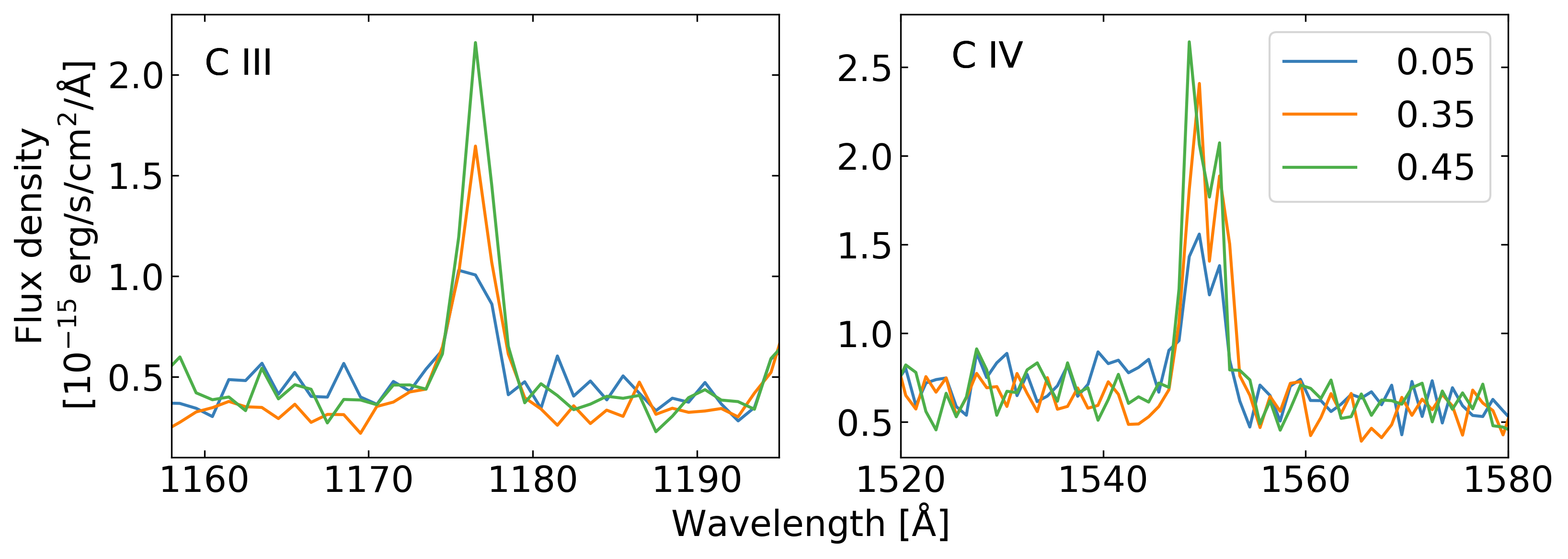}
         \caption{Spectra of two metal emission lines at three phase bins with a width of 0.1 and central value as indicated in the plot label. The emission peaks near phase 0.5 when the irradiated face of the M-dwarf is most visible.
         }
    \label{fig:metals}
\end{figure}

\begin{figure}
	\includegraphics[width=\columnwidth]{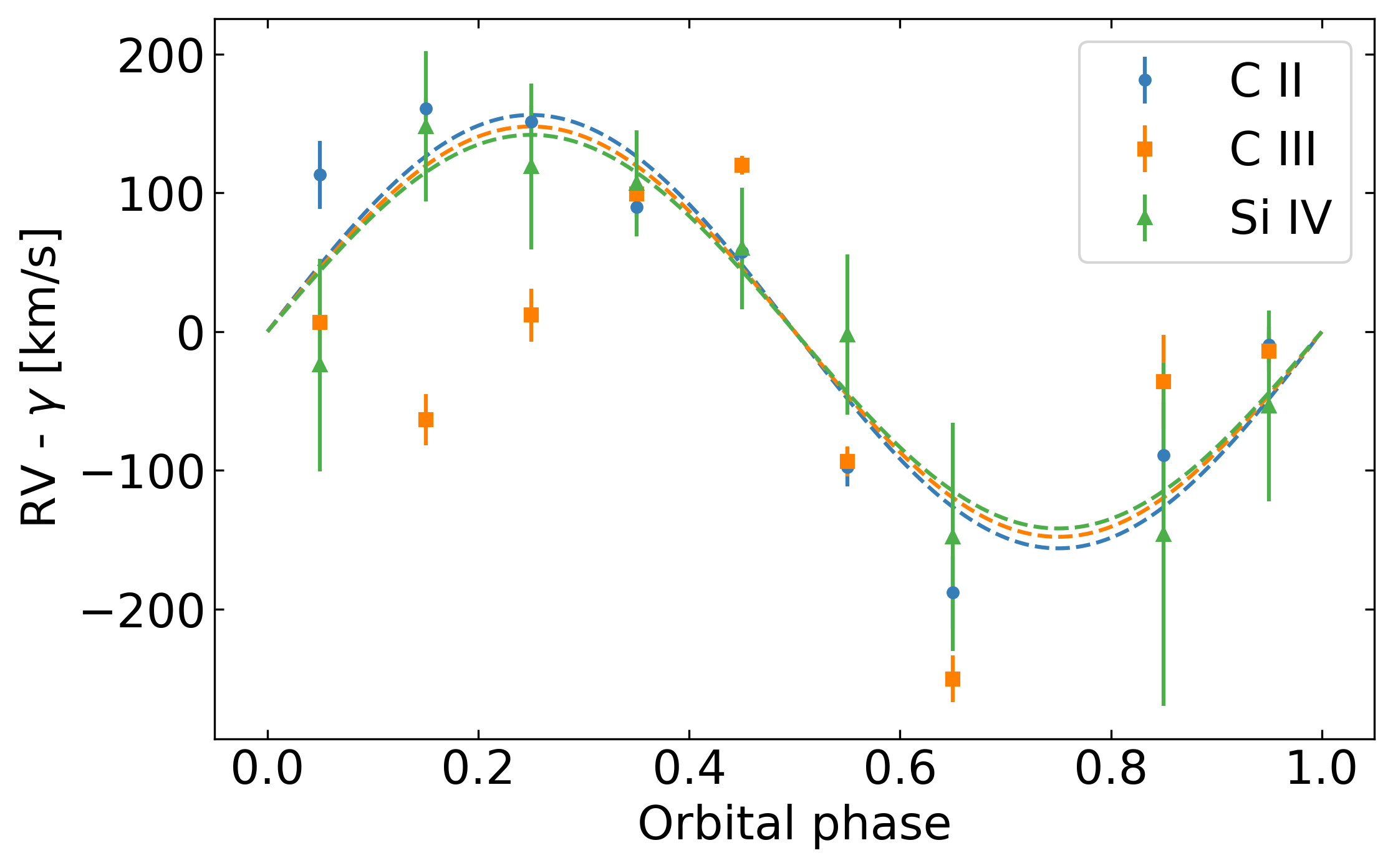}
         \caption{Radial velocities for three metal emission lines visible in \targ's spectrum. The fitted systemic velocity of each line was subtracted because its systematic uncertainty is very large, as the central wavelength of these lines is not precise. 
         }
    \label{fig:rvs}
\end{figure}

The observed metal lines peak in strength at phase 0.5, when the irradiated face of the M-dwarf faces the line of sight (Fig.~\ref{fig:metals}), suggesting that this is where they originate. Using the phase-resolved spectra, we have estimated radial velocities for the C{\sc III} line at $\approx 1175$~\AA, the C{\sc II} line at $\approx 1335$~\AA, and the Si{\sc IV} line at $\approx 1394$~\AA. Other lines are either too weak or blended to yield good results. Radial velocity estimates were carried out by fitting a Gaussian to each line to estimate its observed central wavelength. We then folded the radial velocities using the orbital ephemeris of \citet{Pelisoli2023} and fitted them with $\gamma + K_2\sin(2\pi\varphi)$, where $\gamma$ is the systemic velocity, $K_2$ is the radial velocity semi-amplitude, and $\varphi$ is the orbital phase. The obtained fits are shown in Fig.~\ref{fig:rvs}. The overall behaviour is consistent with these lines tracing the M-dwarf's heated hemisphere, although the C{\sc III} radial velocities show scatter around phase 0.2. We find semi-amplitudes $K_2$ of $138\pm69$, $160\pm27$ and $144\pm18$~km/s for C{\sc III}, C{\sc II} and Si{\sc IV}, respectively. Uncertainties were estimated via bootstrapping.

\begin{figure}
	\includegraphics[width=\columnwidth]{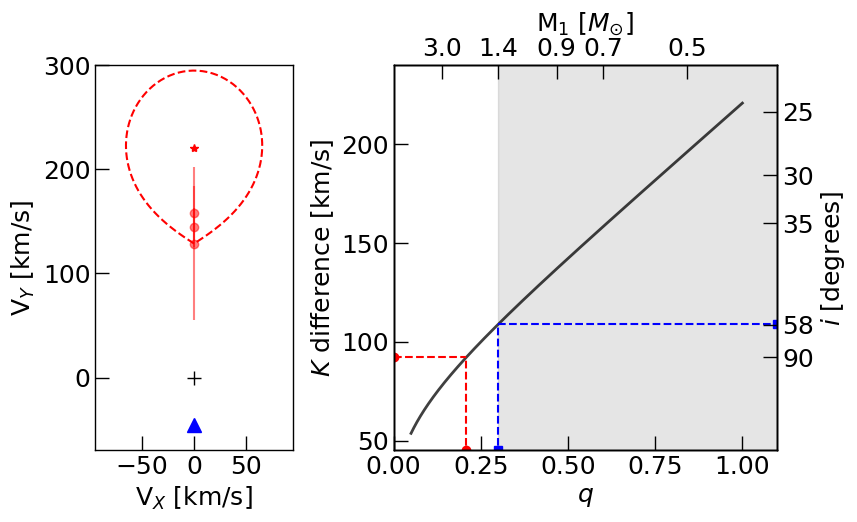}
         \caption{
         The left panel shows different semi-amplitude measurements obtained from phase-resolved spectra. The red star marks the location of the M-dwarf centre of mass according to Na{\sc II} absorption line measurements \citep{Pelisoli2023}. The red circles mark the measurements we derived for C{\sc III}, C{\sc II} and Si{\sc IV}. The red dashed line is the Roche lobe of the M-dwarf for $q = 0.21$, the minimum to encompass the derived semi-amplitudes. The black cross and blue triangle mark the centre of mass of the system and of the white dwarf, respectively. In the right panel, the black line shows the semi-amplitude difference between the M-dwarf centre of mass and its irradiated face, assuming the M-dwarf is Roche-lobe filling. The right-hand y-axis shows the orbital inclination that would correspond to the $K_2$ difference in the left-hand axis. The observed $K_2$ difference sets the minimum $q$ at an inclination of $90^{\circ}$ (red dashed lines). The shaded grey area constraints $M_1$ to white dwarf values, in which case the minimum $q$ is 0.3. Given the semi-amplitude differences measured here, this minimum $q$ corresponds to a maximum inclination of about $58^{\circ}$ (blue dashed lines).
         }
    \label{fig:roche}
\end{figure}

Although the uncertainties are large, the obtained values are systematically smaller than the amplitudes found for \citet{Pelisoli2023} for the Balmer emission lines. This suggests that the metal lines are originated further away from the M-dwarf and closer to the white dwarf and has implications for the Roche constraints. We repeat the same Roche analysis done in \citet{Pelisoli2023} taking the smaller semi-amplitudes derived from the metal lines into consideration. This alone constrains the mass ratio $q = M_2/M_1$ to a minimum value of $q = 0.21$ (left panel in Fig.~\ref{fig:roche}). Requiring the compact object to have a mass lower than the Chandrasekhar limit increases the minimum $q$ to 0.3, and given the $K_2$ difference between the centre of mass and the irradiated face set by the metal emission lines, this implies a maximum orbital inclination of $i \approx 58^{\circ}$ (right panel in Fig.~\ref{fig:roche}).

\section{Geometric model}

\begin{figure}
\includegraphics[width=.99\linewidth]{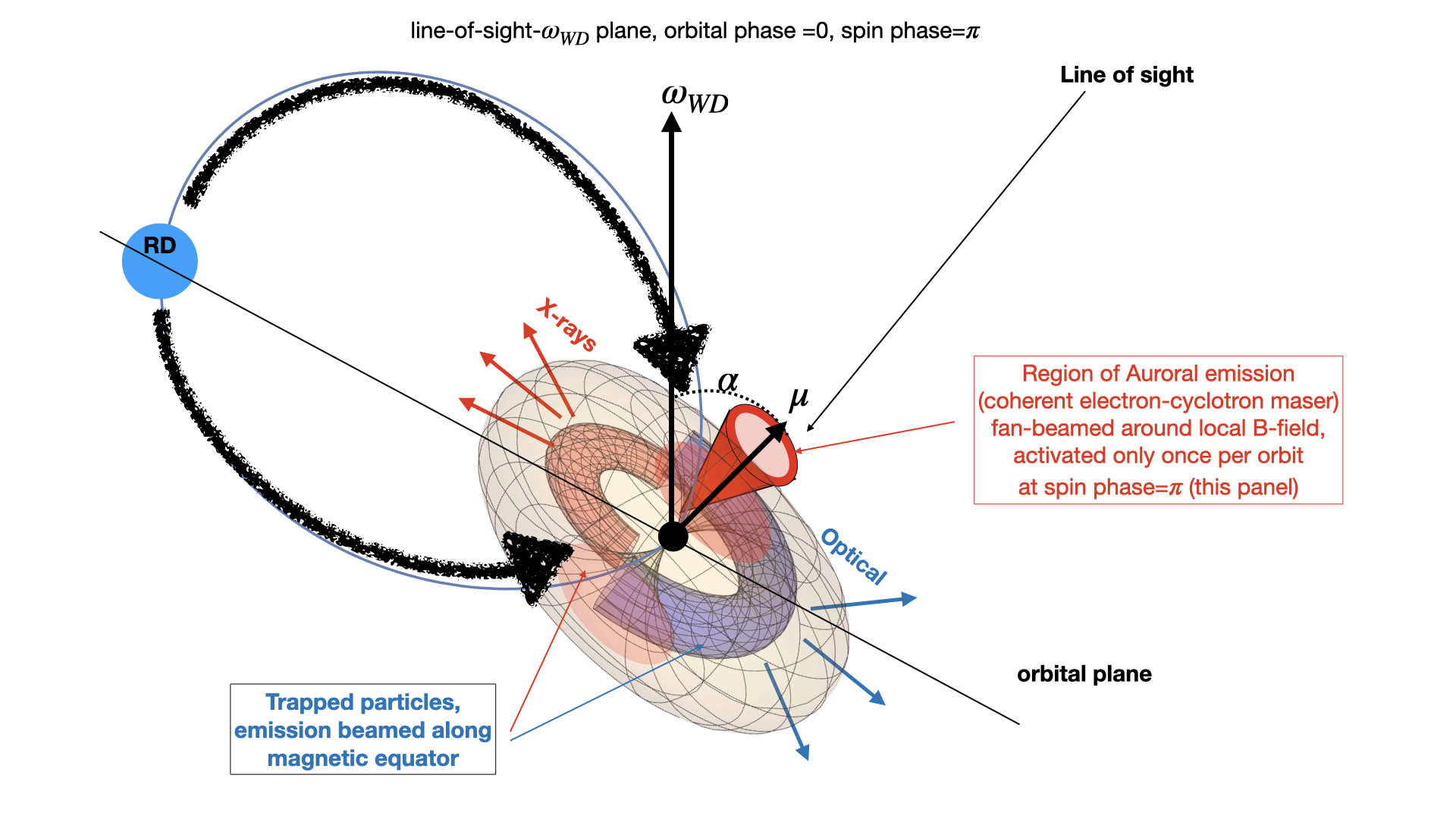}
\caption{Geometry of the system at orbital phase $0$ (superior conjunction of the white dwarf) at the moment when the magnetic moment of the white dwarf ($\mu$) is in the same plane as the white dwarf spin ($\omega_{WD}$) and the line-of-sight plane (spin phase=0.5). A stream from the companion reaches the white dwarf in the magnetic polar region and produces auroral emission (coherent electron-cyclotron maser). Additional emission originates from particles trapped in the Van Allen radiation belts. The resulting emission is mildly beamed along the magnetic equator.
}
\label{overall}
\end{figure}

Based on the observed properties of \targ\ and on the model by \citet{Lyutikov2020}, we propose a geometric model (illustrated in Fig.~\ref{overall}) that can explain many of \targ's characteristics. At the core is the idea that the emission of the white dwarf is seeded by the interaction with the companion. The emission originates in several regions and has different mechanisms.

\begin{figure*}
\includegraphics[width=.49\linewidth]{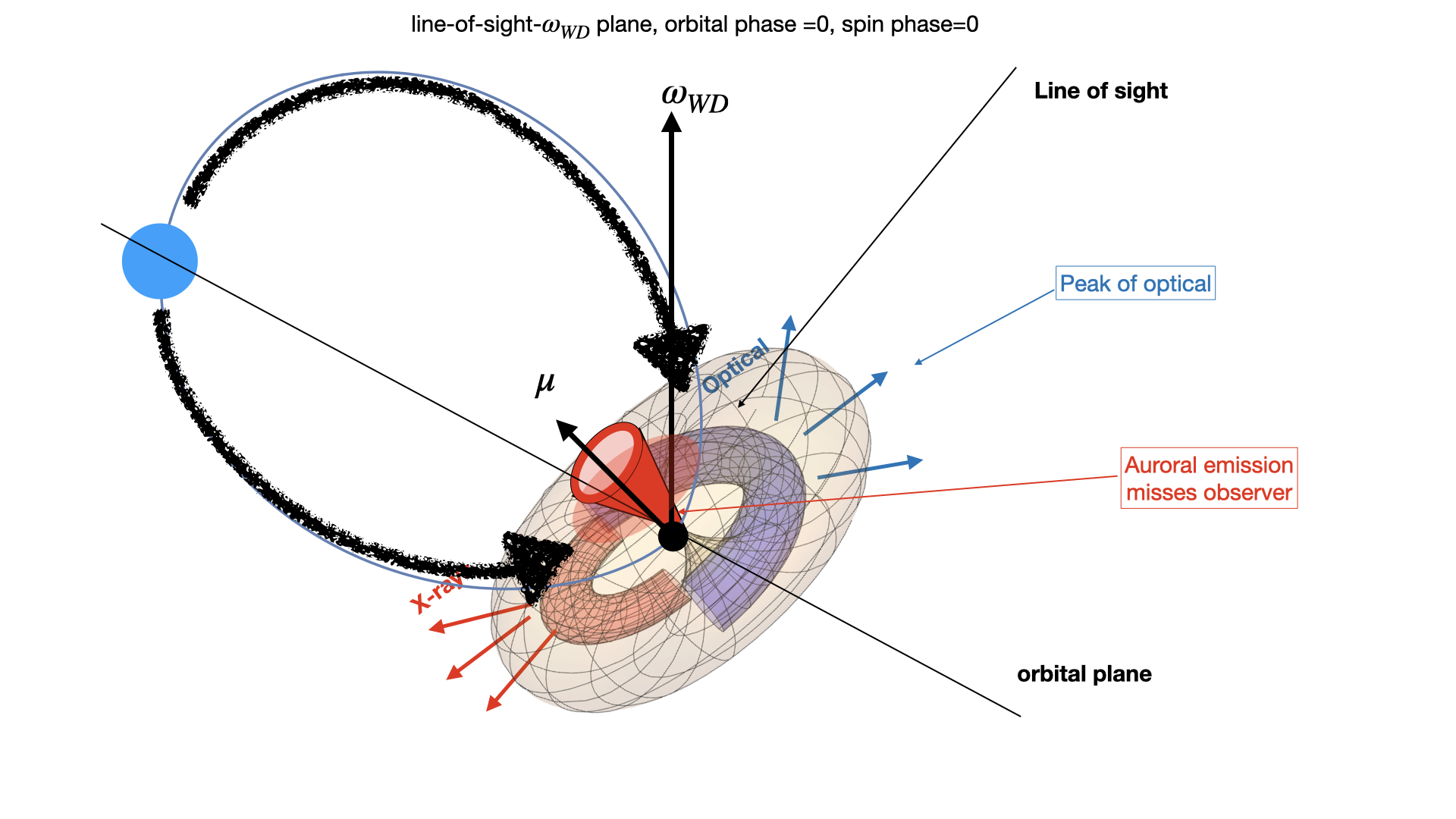}
\includegraphics[width=.49\linewidth]{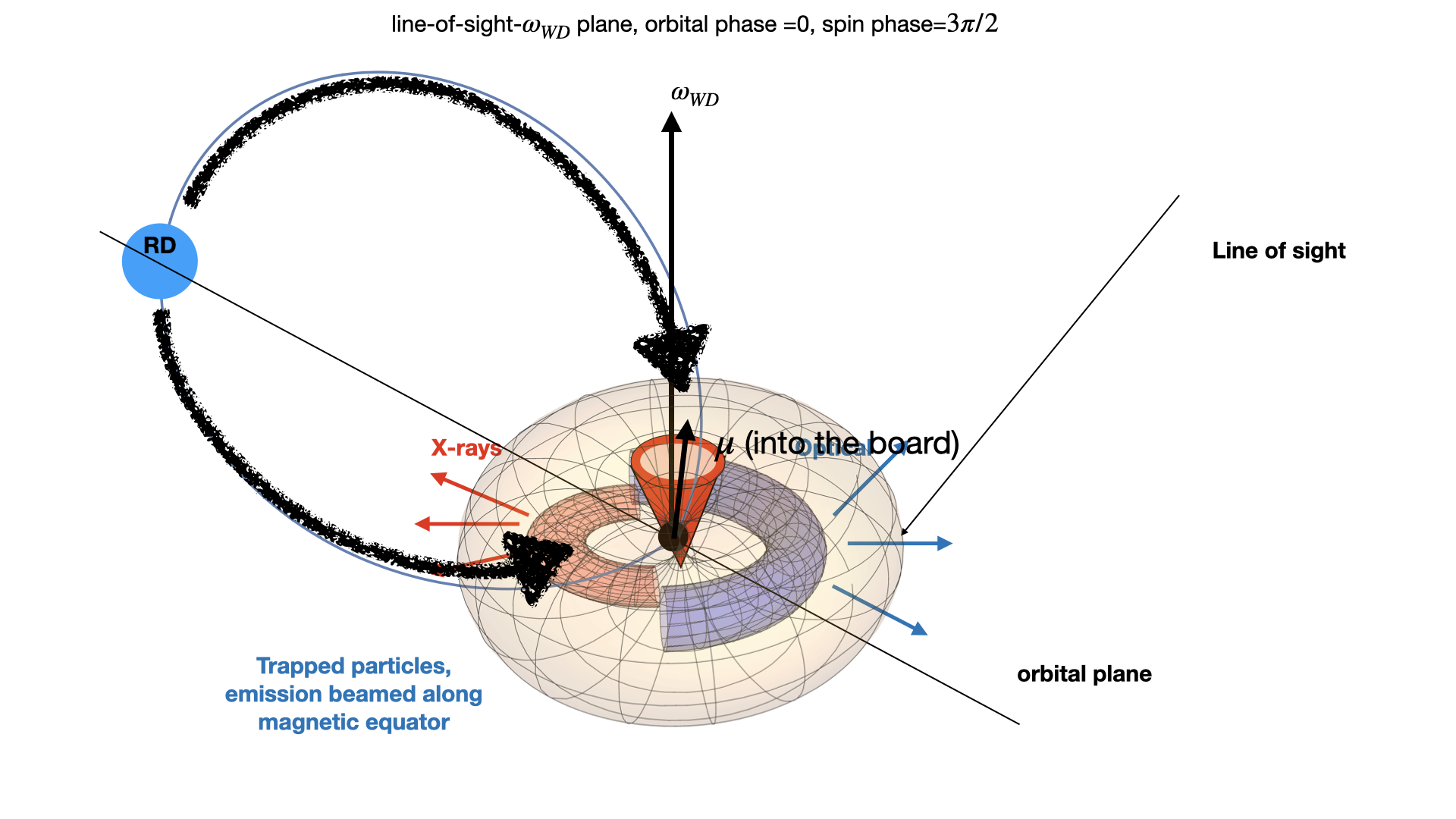}\\
\includegraphics[width=.49\linewidth]{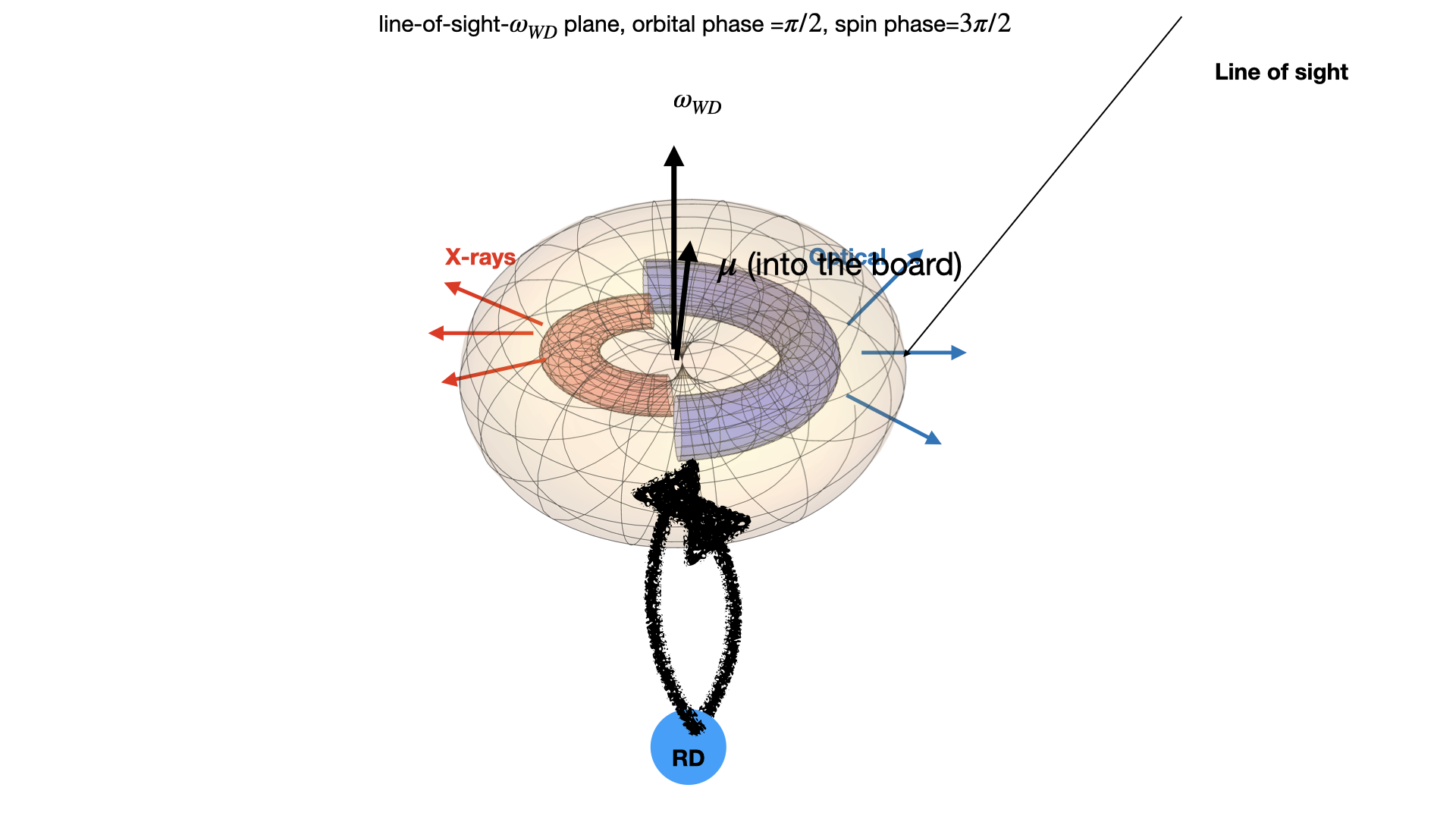}
\includegraphics[width=.49\linewidth]{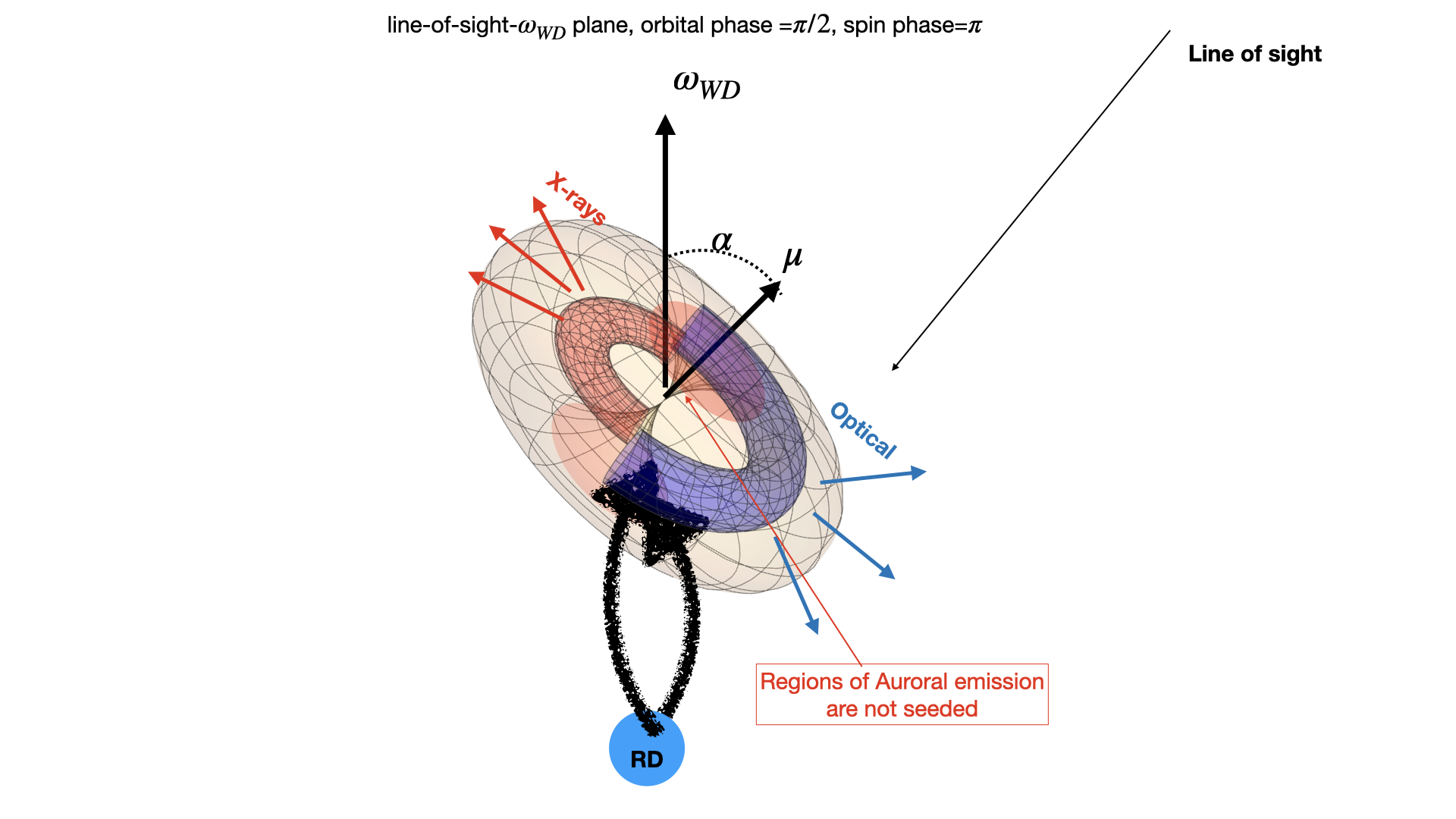}
\caption{Examples of configurations at selected orbital and spin phases. The M-dwarf is represented by the blue circle and seeds particles towards the white dwarf, whose magnetic moment is represented by the red cone and visible only at certain spin phases as indicated. This geometrical model explains the observed behaviour of \targ\ based on the alignment of different emission regions with the line of sight.
}
\label{overall1}
\end{figure*}

We interpret the radio emission as electron-cyclotron maser \citep[e.g.][]{2017RvMPP...1....5M}, operating only in the white dwarf's magnetic polar region and producing emission approximately along the local magnetic field (magnetic moment). This explains one narrow pulse per spin period. In addition, radio emission is activated only when particles from the M-dwarf can reach the polar region. This explains why 
the auroral emission is only observable once per orbit \citep[orbital phase 0.5, see fig. 2 in][]{Pelisoli2023}: particles from the companion can reach the polar regions twice per orbit, at orbital phases $0$ and $0.5$ but, since the emission is highly beamed, it is observed only at orbital phase $0.5$. This geometric configuration is fixed by the fact that radio is seen only at the superior conjunction of the M-dwarf. Activation at the superior or inferior conjunction depends on the relative orientation of the line of sight, the white dwarf's spin and the magnetic moment, see Figs. \ref{overall} and \ref{overall1}. This  geometrical ``seeding" model --  that the emission from the white dwarf is induced by the companion -- works well for explaining the narrow-peaked radio emission.

\begin{figure}
\includegraphics[width=.99\linewidth]{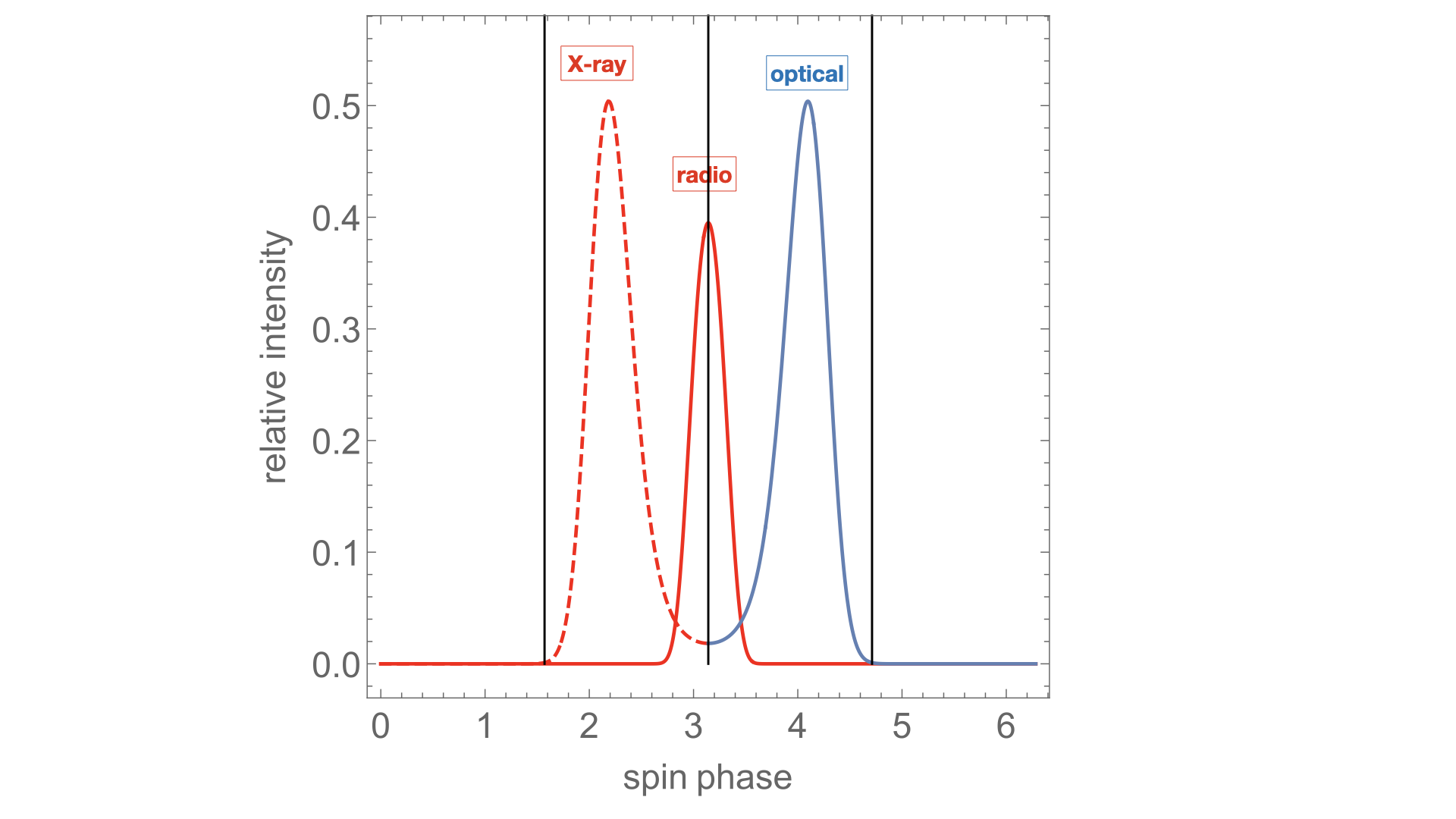}
\caption{Example of theoretical emission profiles, using an arbitrary normalisation. The x-axis shows spin phase in radians. The vertical lines are the spin phases of the maximum of the narrow peak $\pm \pi/2$, to guide the eye. Depending on the parameters of the system, the separation between radio and X-ray/optical peaks may be slightly larger or smaller than $\pi/2$ (0.25) (in this illustrated case, it is smaller).}
\label{profile}
\end{figure}

To model this polar emission we assume that it has a Gaussian profile, so that instantaneous brightness is a Gaussian function of the angle between the line of sight and magnetic moment $\mu$, with a given width \citep[this is a slight simplification, as electron cyclotron maser is expected to have a conical shape around the local magnetic field,][]{2017RvMPP...1....5M}. For the equatorial emission, we assume a Gaussian band along the equator and integrate over equatorial points. This results in the emission shown in Fig.~\ref{profile}.

 The broad X-ray and optical pulses present a challenge -- primarily because of the observed {\it single} peak per spin period. We have investigated two possibilities:
 \begin{itemize} 
\item 
The broad X-ray and optical pulses are due to emissoin by trapped particles in the Van Allen belt, causing beamed emission along the magnetic equator rather than the pole, which would explain the phase shift between the broad and narrow pulses. This X-ray and optical emission is expected with a phase shift of approximately (and not necessarily exactly) $\pm 0.25$ compared to the auroral emission. In fact, the broad pulses can peak around $\pm 0.25$, as can be seen in Fig.~\ref{fig:sphase}.

One important fact remains unexplained by the geometrical model is that it predicts a double peak for the equatorial emission. Instead, X-rays show one single peak (at $-0.25$), while optical shows another single peak (at $+0.25$), with different intrinsic spread. Closer inspection of the ULTRACAM data from \citet{Pelisoli2023} does indicate that {\it occasionally} two shifted components can be seen (Fig.~\ref{fig:peaks}), consistent with this model.

  \begin{figure}
	\includegraphics[width=\columnwidth]{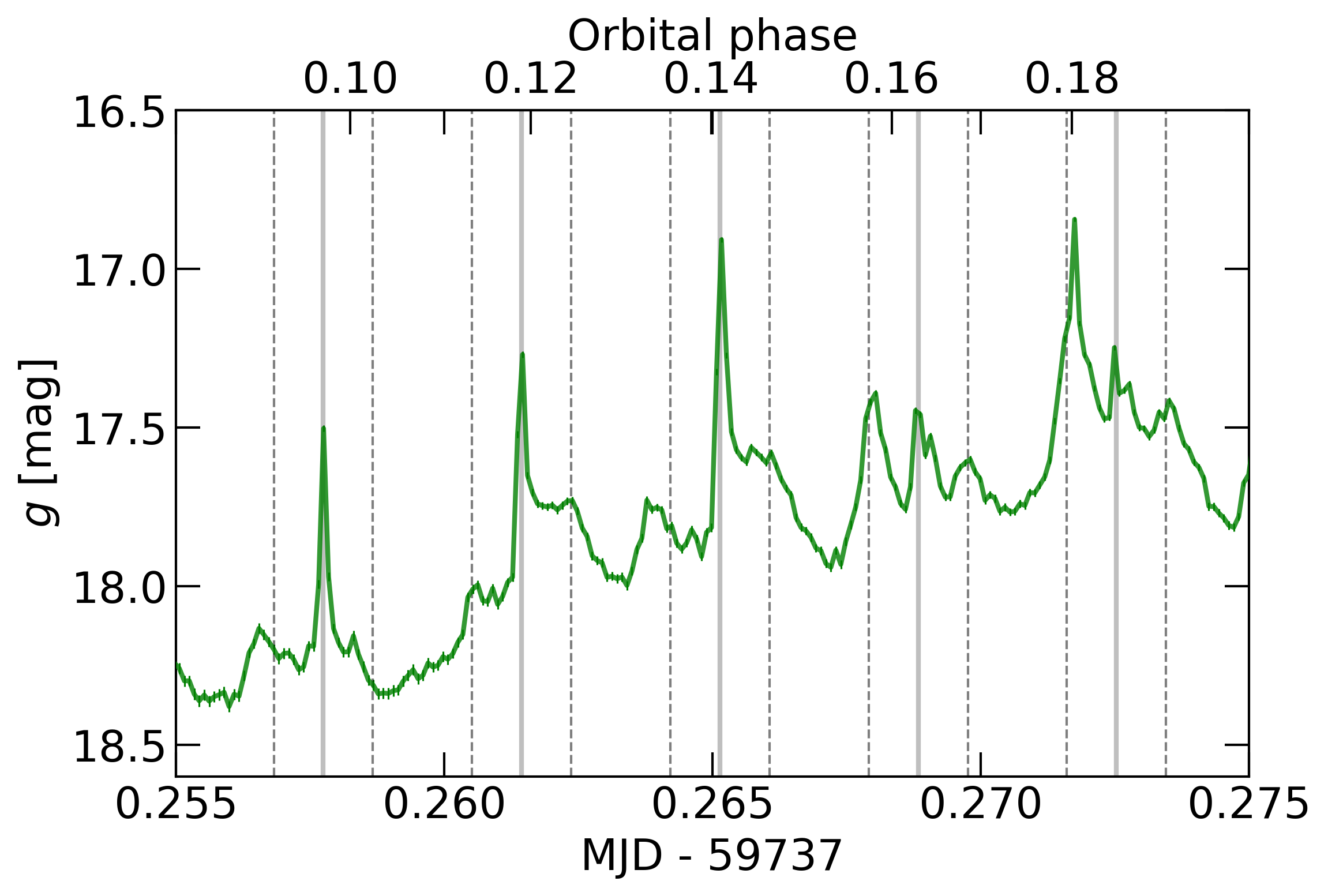}
         \caption{
         Fraction of ULTRACAM data from \citet{Pelisoli2023} taken on 2022 June 06, showing behaviour consistent with the proposed geometric model. The solid vertical lines mark the central peak that could be explained by auroral emission. The dashed lines are displaced by $\pm 0.25$ in the spin phase and coincide with the location of broader peaks that could be due to Van Allen belt emission.
         }
    \label{fig:peaks}
\end{figure}

 \item  Alternatively, the phase-shifted X-ray and optical pulses may come from the plasma stream from the companion interacting with the white dwarf's magnetosphere, as illustrated Fig. \ref{fig:equatorial-injection}. We show theoretical light curves for the following parameters: magnetic obliquity $\alpha=\pi/4$, line of sight with respect to the spin axis $\theta_{ob} = \pi/4$ (near equality of $\alpha $ and $\theta_{ob}$ is needed to see the the narrow polar emission beam in radio), orbital inclination with respect to the white dwarf's spin of $\pi/3$, azimuthal angle of the orbital plane of $\pi/2$ (so that the orbital normal is orthogonal to the line of sight). Different colours correspond to different orbital phases.

  \begin{figure}
	\includegraphics[width=\columnwidth]{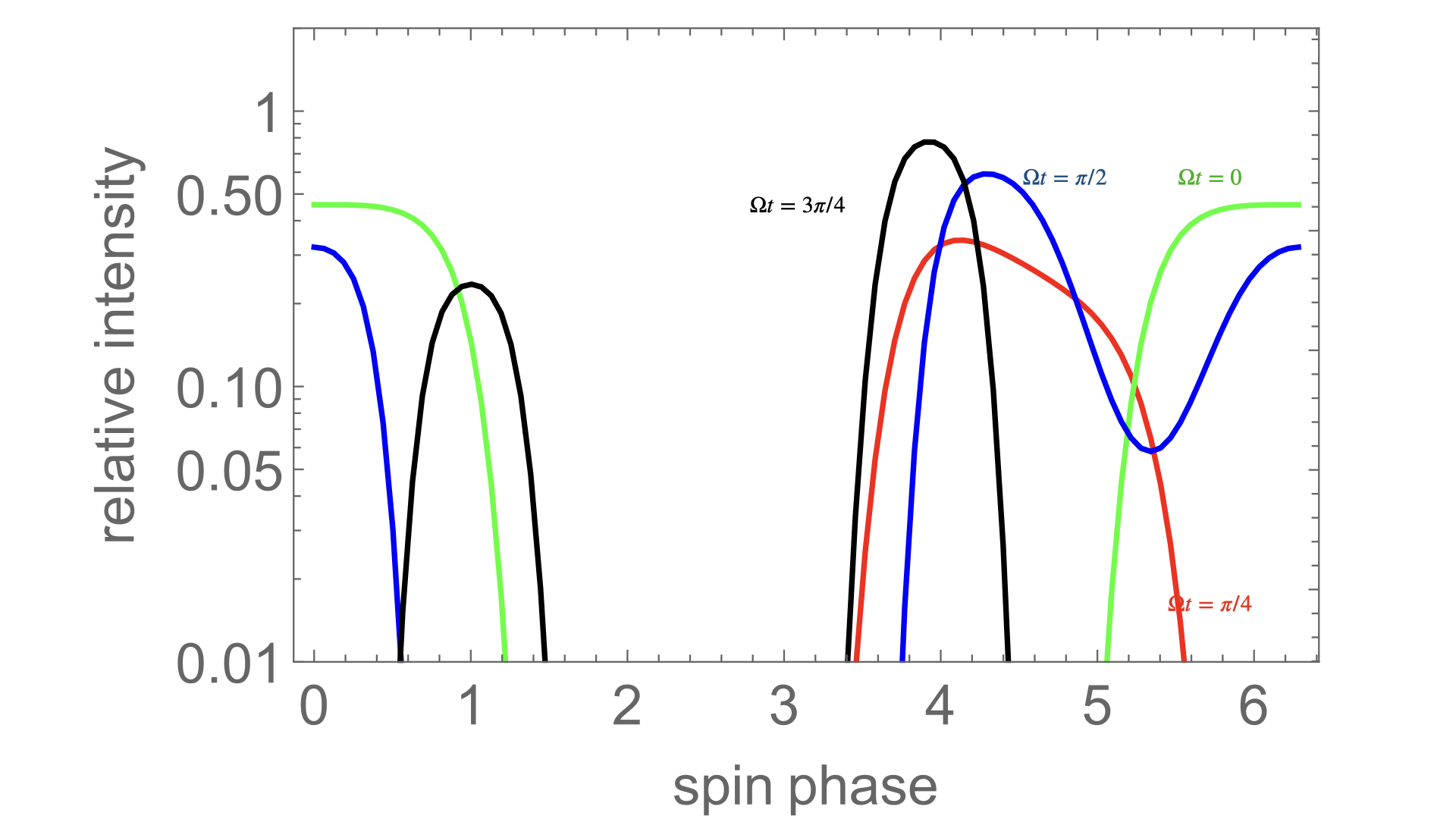}
         \caption{
        Emission profiles expected from the plasma stream from the companion interacting with the white dwarf's magnetosphere in the equatorial plane for orbital phases 0 (green), 0.25 ($\pi/4$, red), 0.5 ($\pi/2$, blue), 0.75 ($3\pi/4$, black).
         }
    \label{fig:equatorial-injection}
\end{figure}

\end{itemize}

 In conclusion, the geometrical model offers ways to understand the single-peaked, and phase-shifted profiles in X-rays and optical. The model generally explains appearances of both  single-peak and double-peak spin profiles. We hypothesise that an important factor is 
 particle diffusion (radial and azimuthal) within the white dwarf's magnetosphere.

Importantly, the model advocates that emission in \targ\ originates/is seeded by the interaction with a companion, which is supported by the fact that such pulsations are not observed in any single white dwarf.

\section{Discussion}

\subsection{The white dwarf in \targ}

The {\it HST} spectra confirm that the compact object in \targ\ is a white dwarf, as previously inferred from its spin period, too slow for a neutron star. The white dwarf temperature obtained from spectral fitting is consistent with the upper limit derived by \citet{Pelisoli2023}. The mass, on the other hand, is $\approx 3\sigma$ below the previously reported value of $1.2\pm0.2$~M$_{\sun}$. To obtain this mass estimate, \citet{Pelisoli2023} relied on the assumption that the semi-amplitudes measured from the Na{\sc ii} doublet and from the H$\beta$ lines traced the centre of mass and the $L1$ Lagrangian point, respectively, with the M-dwarf filling its Roche lobe. This resulted in a maximum inclination estimate of $37^{\circ}$ and on a minimum white dwarf mass of $1.0$~M$_{\sun}$. Although the Na{\sc ii} doublet indeed likely traces the centre of mass, H$\beta$ does not necessarily trace $L1$. In fact, we find that the metal emission lines observed in the FUV suggest a higher minimum inclination of  $58^{\circ}$, which increases the white dwarf minimum mass to $\approx 0.57~M_{\sun}$, consistent with our estimate.

\begin{figure}
	\includegraphics[width=\columnwidth]{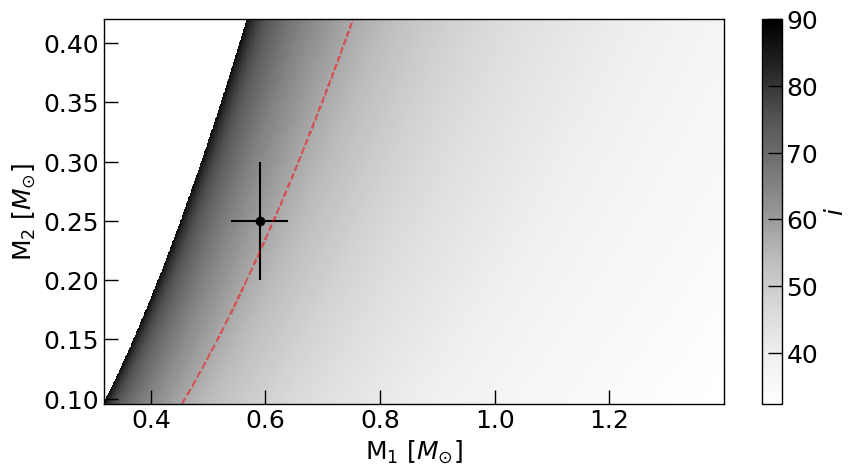}
         \caption{The colour map indicates the system's orbital inclination required to match the observed $K_2$ for the values of $M_1$ and $M_2$ shown in the x- and y-axis. The red dashed line marks the maximum inclination of $58^{\circ}$ inferred from Roche constraints. The black cross shows the mass estimates for white dwarf and M-dwarf from spectroscopic fits, which are completely independent from the Roche analysis. The derived masses are consistent with the Roche constraints.
         }
    \label{fig:masses}
\end{figure}

Fig.~\ref{fig:masses} shows our derived dynamical mass constraints compared to the mass estimates for white dwarf and M-dwarf from spectroscopic fitting. There is good agreement between the minimum values set by a Roche-lobe filling M-dwarf orbiting a white dwarf and the mass estimates. Combining \citet{Pelisoli2023}'s estimate of $0.25\pm0.05$~M$_{\sun}$ for the M-dwarf with the system's binary mass function of $0.1879\pm0.0027$~M$_{\sun}$ and with our estimate for the mass of the white dwarf implies an orbital inclination of $59\pm6^{\circ}$, consistent with the dynamical lower limit.

\subsection{Implications to theoretical models}

Assuming that the white dwarf is in thermal equilibrium, our fit to the white dwarf spectra would suggest that, given the obtained $T\eff$ and mass, crystallisation has not started in the core of the white dwarf \citep{Bedard2020}. This would place a challenge to the rotation- and crystallisation-driven model as a solution for generating a strong magnetic field in the white dwarf, required for explaining the behaviour of binary white dwarf pulsars as a consequence of magnetic torque and reconnection. However, if the white dwarf has previously accreted significantly from the companion, which is probably the case given its spin period, its $T\eff$ will be largely affected by compressional heating, reflecting the past accretion rate rather than the core temperature \citep[e.g][]{Townsley2004, Townsley2009}. The thermal timescale of the white dwarf is $\sim 10^8$~years, which is longer than the binary pulsar phase (as inferred from AR~Sco's spin down). Therefore, the $T\eff$ might not reflect the core temperature of the white dwarf, such that crystallisation remains possible.

Although \targ's $T\eff$ alone does not rule out the occurrence of a rotation- and crystallisation-driven dynamo, it is worth mentioning that there is growing evidence that this model might not be responsible for the late appearance of magnetic fields in white dwarfs. First, for a large fraction of single white dwarfs there does not seem to be a relation between rotation period and field strength \citep[see e.g. fig. 4 in][]{Ginzburg2022}, which is a prediction of the rotation-driven dynamo model. Second, even though there is an observed increase in the relative number of magnetic white dwarfs with decreasing temperature, this actually happens before the onset of crystallisation \citep{Bagnulo2021, Bagnulo2022}. Additionally, numerical simulations by \citet{Fuentes2023} have indicated that the energy released by crystallisation is not sufficient to power a dynamo. Yet, evidence for the late appearance of strong magnetic fields in white dwarfs is overwhelming, both in single and in close binaries systems \citep{Bagnulo2021, Parsons2021, Caron2023, Amorim2023}. If these fields are not triggered by a dynamo, another mechanism must explain their late appearance. Importantly, the evolutionary sequence suggested by \citet{Schreiber2021} does not depend specifically on the crystallisation- and rotation-driven dynamo. It only requires the magnetic field to appear during the cataclysmic variable phase, which could be due to an alternative temperature-dependent mechanism and/or another time-dependent mechanism.

In short, the late appearance of a field remains a possibility for \targ. However, the fact that no magnetic field has been directly detected provides a challenge for models predicting or relying on magnetic fields of the order of $\sim 100$~MG to explain the observed characteristics of pulsing binary white dwarfs \citep[like those by][]{Geng2016, Katz2017, Takata2017}. When the lack of detection was restricted to one system, AR~Sco, perhaps inclination arguments could be used, as that affects the relative strength of different Zeeman components \citep{Honl1925, Unno1956}. However, with two systems displaying the same behaviour and showing no sign of a magnetic field, relying on this argument becomes more challenging. The intensity of the side components varies with $1 + \cos^2(\psi)$, where $\psi$ is the angle between the magnetic field axis and the line of sight. Therefore, both AR~Sco and \targ\ would need to have unfavourable inclinations for the Zeeman side components not to be visible. Additionally, the shape of the Lyman-$\alpha$ line is consistent with the non-magnetic models, which is not what one would expect if a magnetic field were affecting the equivalent width.

These results oppose models requiring a strong magnetic field to explain the behaviour of binary white dwarf pulsars. The alternative models, which require a significant mass-transfer rate, did not seem like a good alternative when only AR~Sco was known, given that it shows no detected flares, but \targ\ has shown evidence for flaring behaviour very similar to propeller systems \citep[see][]{Pelisoli2023}, which could point at a significant mass transfer rate for this system. In fact, we present a geometric model assuming that there is mass transfer that can reproduce most of the characteristics of \targ's light curve.

\section{Summary and conclusions}

We have obtained and analysed {\it HST} COS observations for the recently discovered binary white dwarf pulsar \targ. The data were obtained in TIME-TAG mode, which allowed us to construct a light curve for the observations. The light curve shows strong pulses whose period is consistent with the spin period determined from optical observations. The beat frequency between white dwarf spin and the system's orbital period is detected in the FUV, unlike the optical. The FUV pulse shape is dominated by the broad component which is phase-shifted from the narrow component that is interpreted as tracing the spin of the white dwarf, and hence the FUV pulses cannot be used to improve the spin ephemeris and probe for spin period changes.

We used the light curve to identify times of minima and extract an off-pulse spectrum. Correcting this spectrum for a dilution from the pulse using either the pulse spectrum, a power-law, or constant flux, we obtained estimates for the white dwarf spectrum. Resulting white dwarf parameters were similar for all methods and point at a white dwarf with temperature  $T\eff{}_1 = 11485\pm90$~K and mass  $M_1 = 0.59\pm0.05$~M$_{\sun}$. This suggests that either the white dwarf is not crystallised, or that a previous phase of accretion has led to significant compressional heating such that the white dwarf is not in thermal equilibrium. If the former, this finding would add to growing evidence that crystallisation-driven dynamos are not at play in the core of white dwarfs. If the white dwarf is not in thermal equilibrium, which is possible given that we are observing this system during what is very likely a phase with lifetime shorter than the thermal timescale, it remains possible that the core previously crystallised and that this, combined with the fast rotation, triggered a dynamo. Therefore, it is still possible that the white dwarf was spun up before it became magnetic, as proposed in the evolutionary model of \citet{Schreiber2021}.

We find no evidence for a magnetic field of the order of hundreds of MG, which has implications for proposed models for the observed pulsed emission. The lack of such a detectable magnetic field, combined with the possible detection of flares from \targ, favour models with a significant mass transfer rate to explain the pulses observed in so-called binary white dwarf pulsars, though the absence of detected flaring from AR~Sco remains a puzzle. Continuous monitoring of these systems and searches for other binary radio-pulsing white dwarfs will provide further clues onto the nature of these challenging systems.

\section*{Acknowledgements}

We thank the referee Matthias Schreiber for very helpful scientific insight that improved this manuscript, and Detlev Koester for the use of his atmosphere and spectral codes. 

IP acknowledges support from a Warwick Astrophysics prize post-doctoral fellowship, made possible thanks to a generous philanthropic donation, and thanks the organisers of the workshop Stellar Magnetic Fields from Protostars to Supernovae, held at the Munich Institute for Astro-, Particle and BioPhysics (MIAPbP), which is funded by the Deutsche Forschungsgemeinschaft (DFG, German Research Foundation) under Germany's Excellence Strategy – EXC-2094 – 390783311.

Based on observations made with the NASA/ESA Hubble Space Telescope, obtained at the Space Telescope Science Institute, which is operated by the Association of Universities for Research in Astronomy, Inc., under NASA contract NAS5-26555. These observations are associated with program \#17276.

This project has received funding from the European Research Council (ERC) under the European Union’s Horizon 2020 research and innovation programme (Grant agreement No. 101020057).

This research made use of Astropy (http://www.astropy.org) a community-developed core Python package for Astronomy \cite{astropy:2013, astropy:2018}

For the purpose of open access, the author has applied a Creative Commons Attribution (CC-BY) licence to any Author Accepted Manuscript version arising from this submission.


\section*{Data Availability}

All data analysed in this work can be made available upon reasonable request to the authors.



\bibliographystyle{mnras}
\bibliography{j1912hst} 





\bsp	
\label{lastpage}
\end{document}